\documentclass[12pt,a4paper]{article}
\usepackage[utf8]{inputenc}
\usepackage{a4wide}
\usepackage{graphicx}
\usepackage{amsmath}
\usepackage{amsfonts}
\usepackage{axodraw}
\usepackage{hyperref}
\newcommand\chpt{ChPT}
\newcommand\lag{\ensuremath{\mathcal{L}}}
\newcommand\bO{\ensuremath{\mathcal{O}}}
\newcommand\hPi{\ensuremath{\hat\Pi}}
\newcommand\pip{\ensuremath{{\pi^+}}}
\allowdisplaybreaks[1]

\begin{document}
\begin{titlepage}
\begin{flushright}
LU TP 16-51\\
arXiv:1710.04479 [hep-lat]\\
Revised December 2017
\end{flushright}
\vfill
\begin{center}
{\Large\bf Vector two-point functions in finite volume\\
using partially quenched chiral perturbation theory\\[1.5mm]
at two loops}
\vfill
{\bf Johan Bijnens and Johan Relefors}\\[0.3cm]
{Department of Astronomy and Theoretical Physics, Lund University,\\
S\"olvegatan 14A, SE 223-62 Lund, Sweden}
\end{center}
\vfill
\begin{abstract}
We calculate vector-vector correlation functions at two loops using
partially quenched chiral perturbation theory including finite volume effects
and twisted boundary conditions.
We present expressions for the flavor neutral cases
and the flavor charged case with equal masses.
Using these expressions we give an estimate for the ratio of disconnected to
connected contributions for the strange part of the electromagnetic current.
We give numerical examples for the effects of partial quenching, finite volume
and twisting and suggest the use of different twists to check the size of finite
volume effects.
The main use of this work is expected to be for lattice QCD calculations of
the hadronic vacuum polarization contribution to the muon anomalous magnetic
moment.
\end{abstract}
\vfill
\end{titlepage}
 
\section{Introduction}
The hadronic contribution to the correlation function between two
 electromagnetic currents is known as the hadronic vacuum polarization (HVP).
 An important application of the HVP is in the prediction of the anomalous
 magnetic moment of the muon, muon $g-2$. The muon $g-2$ is defined by
\begin{align}
  a_\mu = \frac{g_\mu-2}{2}
\end{align}
where $g_\mu$, the gyromagnetic ratio, is one of the best measured quantities
 in physics. The experimental value from \cite{gm2exp1,gm2exp2,gm2exp3,PDG} is
\begin{align}
  a_\mu = 11659208.9(5.4)(3.3) 10^{-10}.
\end{align}
This value is 3 to 4 standard deviations away from the standard model (SM)
 prediction, where
 the precise tension depends on which prediction is used,
 see \cite{Jegerlehner:2009ry} for a review and \cite{Proceedings:2016bng}
 for more recent discussions. A new experiment at Fermilab aims to improve
 the uncertainty in the experimental measurement
 to $0.14$ ppm \cite{gm2expFNAL} and there are even more ambitious reductions
 in the uncertainty discussed in \cite{gm2expJPARC}. However, in order to take
 full advantage of the reduced experimental errors the theoretical prediction
 must also be improved.

The theoretical prediction is usually divided into a pure QED, an electroweak
 and a hadronic contribution
\begin{align}
  a_\mu = a_\mu^{\textrm{QED}} + a_\mu^{\textrm{EW}} + a_\mu^{\textrm{had}}.
\end{align}
The main uncertainty in current predictions come from the hadronic part. This
 part can be divided into lowest order, higher orders and light-by-light
 contributions;
\begin{align}
  a^{\textrm{had}}_\mu = a_\mu^{\textrm{LO-HVP}} + a_\mu^{\textrm{HO-HVP}} + a_\mu^{\textrm{HLbL}}.
\end{align}
The first and last term dominate the uncertainty. For a nice overview of the
 different contributions and their uncertainties,
 see Fig. 19 in \cite{Gorringe:2015cma}. In the following we focus on the
 first term which is related to the HVP.

$a_\mu^{\textrm{LO-HVP}}$ can be determined in several ways. One way is to use
 dispersion relations to relate
$a_\mu^{\textrm{LO-HVP}}$ to $\sigma(e^+e^- \rightarrow \textrm{hadrons})$ or
 $\sigma(\tau\rightarrow \nu_\tau + \textrm{hadrons})$. There is some tension
 between the two determinations \cite{PDG}. This highlights the need for other
 ways of determining the HVP contribution to the muon $g-2$. One possibility
 is using lattice QCD\footnote{A recent proposal on the experimental side is
 given in \cite{Calame:2015fva}.}.

In lattice QCD, the HVP 
is evaluated at Euclidean momentum transfer \cite{Blum:2002ii}. A complication
 is that the most important contributions to $a_\mu^{\textrm{LO-HVP}}$ are
 with Euclidean $Q^2 \simeq m_\mu^2 \simeq (106~\textrm{MeV})^2$. The
 contributions from different momentum regions are discussed in
 Fig. 3 of~\cite{DellaMorte:2011aa}. Simulating with periodic boundary
 conditions around $Q^2 \simeq m_\mu^2$ would require much larger volumes
 than presently available and there are also complications
 around $Q^2 \simeq 0$.

There are a number of proposals how these difficulties can be overcome. The
 use of partially twisted boundary conditions to allow continuous variation
 of momenta was given in \cite{DellaMorte:2010aq,Juttner:2009yb}, see
 also \cite{Aubin:2013daa}. This is only possible for the connected parts
 of the HVP and there is an added complication in that the cubic symmetry
 of the lattice is further
 reduced \cite{Bernecker:2011gh,Blum:2016xpd,Bijnens:2014yya}. Some other
 recent proposals and calculations are given
 in \cite{Bali:2015msa,Feng:2013xsa,Chakraborty:2014mwa,deRafael:2014gxa,Dominguez:2016eol,Bodenstein:2013flq,Golterman:2014ksa}. 
The present status of lattice QCD determinations of hadronic contributions
 to the muon $g-2$ was outlined in \cite{WittigLattice2016}.

In this paper we focus on effects from finite volume, partially twisted
 boundary conditions and partial quenching (PQ) using PQ chiral chiral
 perturbation theory (PQ\chpt{}). Finite volume effects for the HVP were
 studied in \cite{Aubin:2015rzx} where they found that chiral perturbation
 theory (\chpt) gives a good description of the finite volume effect already
 at leading order, which is $p^4$ in this case. Finite volume corrections
 using a different approach have been discussed in \cite{Francis:2013qna}.

Here we calculate general vector two-point functions in PQ\chpt{} in finite
 volume, that is both the finite volume correction and the infinite volume
 part, with twisted boundary conditions at $p^6$. Previous results
 in \chpt\ with twisted boundary conditions at $p^4$ were given
 in \cite{Bijnens:2014yya,Aubin:2015rzx}. We also point out that the
 finite volume corrections may be estimated by using different twist
 angles at the same $q^2$ in the same ensemble. Note that we use Minkowski
 space conventions.

In \cite{DellaMorte:2010aq,Francis:2013qna,Bijnens:2016ndo} the ratio of disconnected to
 connected contributions for various contributions to the HVP were discussed.
Here we extend the order $p^6$ analysis of \cite{Bijnens:2016ndo} 
to the ratio for the strange quark contribution to the electromagnetic current.
We use the assumption of vector meson dominance (VMD) for the $\phi$ meson
(VMD$\phi$) for the pure low-energy-constant (LEC) contribution in PQ\chpt. 

This paper is organized as follows. In section \ref{sec:VVcorr} we introduce
 the vector two-point function in finite volume with twisted boundary
 conditions. Section \ref{sec:chpt} gives a brief introduction to
 PQ\chpt\ with twisted boundary conditions. Our main results, the
 expressions for the one-point and two-point functions to order $p^6$ in
 PQ\chpt\ are introduced in section \ref{sec:analytical}.
There we also present the $p^4$ expressions. The expressions at $p^6$ are
 given in the appendix where the integral notation used is also introduced.
 In section \ref{sec:discon} we discuss the ratio of disconnected to connected
 contributions in PQ\chpt, extending the analyses in \cite{Bijnens:2016ndo}.
 In section \ref{disconnectedstrange} we estimate the ratio of disconnected
 to connected contributions to the strange part of the electromagnetic current.
 We then present some numerical examples and a way to estimate finite volume
 effects using lattice data in section \ref{sec:numerics}. Finally we conclude
 in section \ref{sec:conclusion}. 

An earlier version of this paper appeared in \cite{thesisjohan}. The numerical
programs will be made available
in \textsc{CHIRON} \cite{Bijnens:2014gsa,chiron}.

\section{VV correlation function}
\label{sec:VVcorr}

We define the vector two-point function as
\begin{align}
  \Pi^{\mu\nu}_{ab}(q) =
 i\int d^4x \exp(iq\cdot x)\left<T\left(j^\mu_a(x)j^{\nu\dagger}_b(0)\right)\right>
\end{align}
with $a,b$ indicating which currents are being considered. In cases
 where $a=b$ we use 
\begin{align}
  \Pi^{\mu\nu}_{a}(q) \equiv  \Pi^{\mu\nu}_{ab}(q),\quad a=b.
\end{align}

We define the electromagnetic current as
\begin{align}
  j^\mu_{EM} = \frac{2}{3}j^\mu_U-\frac{1}{3}j^\mu_D-\frac{1}{3}j^\mu_S
\end{align}
where
\begin{align}
  j^\mu_U = \bar u\gamma^\mu u,\qquad j^\mu_D = \bar d\gamma^\mu d,\qquad j^\mu_S = \bar s\gamma^\mu s.
\end{align}
In order to be able to apply twisted boundary conditions for the connected
 part of various two-point functions we will also define the off diagonal
 vector current
\begin{align}
  j^\mu_{\pi^+_v} = \bar d\gamma^\mu u.
\end{align}

The combination of two electromagnetic currents can be written as
\begin{align}
  j^\mu_{EM}j^{\nu\dagger}_{EM} = \frac{1}{9}\left(4j^{\mu}_Uj^{\nu\dagger}_U + j^{\mu}_Dj^{\nu\dagger}_D + j^{\mu}_Sj^{\nu\dagger}_S - 4j^{\mu}_Uj^{\nu\dagger}_D - 4j^{\mu}_Uj^{\nu\dagger}_S + 2j^{\mu}_Dj^{\nu\dagger}_S \right).
\end{align}
We do not consider the corresponding two-point functions one by one. Instead
 we use the fact that in PQ\chpt{} we can keep the masses of the valence
 quarks arbitrary and calculate only one connected and one disconnected
 two-point function. We denote these by
\begin{align}
  \label{eq:toPointFcns}
  \Pi^{\mu\nu}_{\pi^+_v}\qquad\textrm{and}\qquad \Pi^{\mu\nu}_{XY},
\end{align}
where $X,Y\in U,D,S$ with $X\neq Y$. These can then be used to construct all
 the possible two-point functions. The finite volume correction for the
 connected parts of $\Pi^{\mu\nu}_{EM}(q)$ calculated at arbitrary momentum 
transfer using twisted boundary conditions can be estimated
 from $\Pi^{\mu\nu}_{\pi^+_v}$. As it stands, $\Pi^{\mu\nu}_{\pi^+_v}$ is
 related to the connected part of $\Pi^{\mu\nu}_{U}$ but, setting the up and
 down valence quark masses to the strange quark mass, the connected part
 of $\Pi^{\mu\nu}_{S}$ can also be accessed. In this way the expressions are
 more general than the notation might imply. This is enough for calculating
 the connected part of the HVP with twisted boundary conditions. 

There are constraints on the form factors following from the Ward identity
\begin{align}
  \partial_\mu\bar q_i\gamma^\mu q_j = i\left(m_i - m_j\right)\bar q_i q_j.
\end{align}
We only consider currents with same-mass quarks in which case the right hand side is zero and the current is conserved. In infinite volume this leads to the relation
\begin{align}
  \label{eq:wardInf}
  \partial_\mu\Pi^{\mu\nu}_{ab} = 0.
\end{align}
For the case of the electromagnetic current this also follows from gauge invariance. In a Lorentz invariant framework any two-point function constructed from conserved currents can be written as
\begin{align}
  \label{eq:lorentzWard}
  \Pi^{\mu\nu}_{ab} = \left(q^\mu q^\nu - q^2g^{\mu\nu}\right)\Pi_{ab}(q^2).
\end{align}
The quantity which is needed for the calculation of the muon $g-2$ is the
subtracted quantity
\begin{align}
  \hPi_{ab}(q^2) = \Pi_{ab}(q^2) - \Pi_{ab}(0)
\end{align}
where $a=b=EM$.

In finite volume, (\ref{eq:wardInf}) doesn't hold for off-diagonal currents. In this case we get instead
\begin{align}
  \label{eq:wardFin}
  i\partial_\mu\left<T\left\{j_{\pi^+_v}^\mu(x) j_{\pi^+_v}^{\nu\dagger}(0)\right\}\right> = \delta^{(4)}(x)\left<\bar d \gamma^\nu d - \bar u\gamma^\nu u\right>.
\end{align}
The right hand side contains vacuum expectation values (VEVs) of flavor neutral vector currents which can be non-zero due to broken Lorentz symmetry.
In particular, different twists for the up and down quarks will make the right hand side in (\ref{eq:wardFin}) non-zero.
Broken Lorentz symmetry also means that the decomposition (\ref{eq:lorentzWard}) can not be used. In our results we use the parameterization (note that $\Pi_{1ab}$ has no factor of $q^2$ in front)
\begin{align}
  \label{eq:FVsplit}
  \Pi^{\mu\nu}_{ab} = q^\mu q^\nu\Pi_{0ab}(q) - g^{\mu\nu}\Pi_{1ab}(q) + \Pi^{\mu\nu}_{2ab}(q).
\end{align}
This split is not unique but provides a useful way to organize results. Expressions given in this form reduce to (\ref{eq:lorentzWard}) in the infinite volume limit. The Ward identity for $\Pi^{\mu\nu}_{\pi^+_v}$ following from (\ref{eq:wardFin}) is
\begin{align}
  \label{eq:wardSplitPip}
  q^2 q^\nu\Pi_{0\pi^+_v}(q) - q^\nu\Pi_{1\pi^+_v}(q) + q_\mu\Pi^{\mu\nu}_{2\pi^+_v}(q) = \left<\bar u \gamma^\nu u - \bar d\gamma^\nu d\right>.
\end{align}
For $\Pi^{\mu\nu}_{XY}$ we obtain instead
\begin{align}
  \label{eq:wardSplitXY}
  q^2 q^\nu\Pi_{0XY}(q) - q^\nu\Pi_{1XY}(q) + q_\mu\Pi^{\mu\nu}_{2XY}(q) = 0.
\end{align}
We have used these Ward identities to verify both our analytical expressions and numerical programs.

\section{Partially quenched \chpt{} and twisted boundary conditions}
\label{sec:chpt}

The low energy effective field theory for the lightest pseudoscalar mesons is \chpt{} \cite{Weinberg1,GL1,GL2}. One way to parameterize the mesons in \chpt{} is
\begin{align}
  U = \exp\left(i\sqrt 2 \frac{M}{F_0}\right),\,
  M=\left(
    \begin{matrix}
      \frac{\pi^0}{\sqrt 2} + \frac{\eta}{\sqrt 6} & \pi^+ & K^+\\
      \pi^- & -\frac{\pi^0}{\sqrt 2} + \frac{\eta}{\sqrt 6} & K^0\\
      K^- & \bar K^0 & -\frac{2\eta}{\sqrt 6}
    \end{matrix}
  \right),
\end{align}
where $F_0$ is the pion decay constant in the chiral limit. The trace of $M$, corresponding to the singlet $\eta$, is removed due to the anomaly. To include partial quenching in \chpt{} we keep the trace of $M$ and include a mass term for the singlet $\eta$ which can be sent to infinity at a later stage \cite{Sharpe:2001fh}.

$M$ is then redefined as
\begin{align}
  M=\left(
    \begin{matrix}
      U & \pi^+ & K^+\\
      \pi^- & D & K^0\\
      K^- & \bar K^0 & S
    \end{matrix}
  \right),
\end{align}
where $U,D,S$ are flavor neutral mesons with quark content $\bar u u,\bar d d,\bar s s$ respectively. It is then possible to interpret the indices of $M$ as flavor indices. Flavor indices can then be followed in Feynman diagrams using a double line notation for the mesons. Flavor lines forming loops are summed over all flavors and correspond to sea flavors, and lines which are connected with external mesons have fixed flavor content corresponding to valence flavors. Setting the masses of mesons with valence-valence, sea-valence or sea-sea meson different incorporates partial quenching. The method of following flavor lines is known as the quark flow method \cite{Sharpe:1990me,Sharpe:1992ft,Bernard:1992mk}.

The lowest order Lagrangian with a singlet $\eta$ mass term is
\begin{align}
  \lag = \frac{F_0^2}{4}\left<D_\mu U D^\mu U^\dagger\right> + \frac{F_0^2}{4}\left<\chi U^\dagger + U\chi^\dagger\right> + \frac{m_0^2}{3}\left(U+D+S\right)^2,
\end{align}
where $\left<\ldots\right>$ denotes the trace of $\ldots$ in flavor space and
\begin{align}
  D_\mu U =\,& \partial_\mu U - i r_\mu U + i U l_\mu,&
  \chi =\,& 2B_0(s+ip),
\end{align}
with $r_\mu,l_\mu,s,p$ external fields or sources. $F_0$ is the pion decay constant in the chiral limit and  $B_0$ is related to the scalar quark condensate. The external sources will be used for incorporating quark masses, interactions with external photons and to generate Green functions of all our two-point functions.

Quark masses are included by setting
\begin{align}
  s =
  \left(\begin{matrix}
      m_u & 0 & 0\\
      0 & m_d & 0\\
      0 & 0 & m_s
    \end{matrix}\right),
\end{align}
where valence masses should be used for a fixed index on $s$ and sea masses should be used for a summed index on $s$. External photons are introduced by
\begin{align}
  v_\mu = l_\mu = r_\mu = eA_\mu\left(\begin{matrix}
      2/3 & 0 & 0\\
      0 & -1/3 & 0\\
      0 & 0 & -1/3
    \end{matrix}\right),
\end{align}
where $A_\mu$ is the external photon field and $e$ is the electromagnetic charge.

In order to calculate two-point functions such as $\Pi_{UU}$, we need to use 
\begin{align}
  v_\mu = V_\mu\left(\begin{matrix}
      1 & 0 & 0\\
      0 & 0 & 0\\
      0 & 0 & 0
    \end{matrix}\right),
\end{align}
where $V_\mu$ is an external vector field. The standard \chpt{} Lagrangian assumes that the matrix $v_\mu$ is traceless which is not the case here. Including the trace of $v_\mu$ leads to additional terms in the Lagrangian. As explored in Ref. \cite{Bijnens:2016ndo} these extra terms do not couple to mesons until $\bO(p^6)$, or $\bO(p^4)$ via the Wess-Zumino-Witten (WZW) term. For the two-point function, two such vertices are needed. There is then no contribution to the finite volume correction until $\bO(p^8)$. The $\bO(p^6)$ terms do influence the infinite volume expressions and are needed in order to render these finite. The $\bO(p^4)$ and $\bO(p^6)$ Lagrangians can be found in \cite{GL1,GL2} and \cite{BCE1,BCE2}, respectively.

The main extra complication from the singlet $\eta$ mass term is that the propagator for diagonal mesons becomes rather involved. After the limit $m_0\rightarrow\infty$ is taken the propagator between an $a\bar a$ and $b\bar b$ meson is 
\begin{align}
  \label{eq:diagProp}
  G_{ab}=& \frac{i\delta_{ab}}{p^2-m_a^2} +i \mathcal{D}_{ab},\nonumber\\
  \mathcal{D}_{ab} =&  - \frac{1}{3}\frac{(p^2-m_1^2)(p^2-m_2^2)(p^2-m_3^2)}{(p^2-m_a^2)(p^2-m_b^2)(p^2-m_{\pi^0}^2)(p^2-m_{\eta}^2)},
\end{align}
where $m_{1,2,3}$ are sea quark masses. For numerical integration we evaluate integrals with this propagator using the residue notation given in \cite{Aubin:2003mg}. However, in the analytical expressions we keep $\mathcal{D}_{ab}$ intact, see Appendix \ref{app:integrals}.

For a quark $q$ in a box with length $L$, twisted boundary conditions are defined by
\begin{align}
  q(x^i + L) = \exp(i\theta^i_q)q(x^i),
\end{align}
where $\theta^i_q$ is the twist angle in the $i$ direction. The twist of the anti-quark follows from complex conjugation. The allowed momenta in direction $i$ of the quark are then
\begin{align}
  p^i = \frac{2\pi}{L}n + \frac{\theta^i_q}{L},\qquad n\in\mathbb{Z}.
\end{align}
The momentum of the quark can be continuously varied by varying the twist angle.

In \cite{Sachrajda:2004mi}, \chpt\ with twisted and partially twisted boundary conditions was developed, where partial twisting means that the twist on valence and sea quarks are different. The twist of a $\bar q^\prime q$ meson is
\begin{align}
\label{eq:twistmeson}
  \phi_{\bar q^\prime q}(x^i+L) = \exp(i(\theta^i_q-\theta^i_{q^\prime})) \phi_{\bar q^\prime q}(x^i).
\end{align}
Diagonal mesons have zero twist and charge conjugate mesons have opposite twists of one another.

Loop integrals are replaced by sums over allowed momenta in finite volume. We regulate our integrals using dimensional regularization giving that we replace
\begin{align}
  \label{eq:twistInt}
  \int \frac{d^d k}{(2\pi)^d} \rightarrow \int_V\frac{d^d k}{(2\pi)^d} = \int \frac{d^{d-3} k}{(2\pi)^{d-3}} \sum_{\vec k = \frac{2\pi}{L}\vec n + \frac{\vec \theta}{L}}
\end{align}
where we have collected the twist angles $\theta^i$ in a vector $\vec \theta$. We also use the four vector notation $\theta^\mu = (0,\vec \theta)$.
The angles $\theta^i$ are derived from (\ref{eq:twistmeson})
for a meson with flavour content $\bar q^\prime q$ travelling in the loop
and are $\theta^i= \theta^-_q-\theta^i_{q^\prime}$.

An important consequence of twisted boundary conditions is that the summation in (\ref{eq:twistInt}) is not symmetric around zero, which gives
\begin{align}
  \int_V\frac{d^d k}{(2\pi)^d}\frac{k^\mu}{k^2-m^2}\neq 0.
\end{align}
This is a consequence of the fact that twisted boundary conditions break the cubic symmetry of the lattice. The way we evaluate integrals in finite volume is described in Appendix~\ref{app:integrals}.

\section{Analytical results}
\label{sec:analytical}

In this section we give expressions for the vector one-point and two-point functions at $p^4$. The expressions at $p^6$ are given in Appendix \ref{app:results} since they are rather long. We denote the $p^4$ part of a quantity $X$ by $X^{(4)}$ and the $p^6$ part is denoted by $X^{(6)}$. Note that the results contain implicit sums over sea quarks. A term containing both $\mathcal{S}$ and $\mathcal{S}^\prime$ has two implicit sums, a term containing only $\mathcal{S}$ has one implicit sum and a term with no sea quark mentioned has no implicit sum. 

The results in Appendix \ref{app:results} contain both the finite volume correction and the infinite volume part. For a quantity $X$ we denote this by $X^{\mathcal{V}}$. If we would write these out separately the infinite volume part would be a bit shorter but the finite volume correction would be much longer. To achieve this compact expression we write every integral in finite volume as the sum of the finite part of the infinite volume integral after renormalization plus the finite volume correction. Symbolically we use the notation where the part of an integral $A$ which remains after renormalization is written as
\begin{align}
  \label{eq:mathcalV}
  A^{\mathcal{V}} = \bar A + A^V.
\end{align}
This is described in more detail in Appendix \ref{app:integrals}.
Note that for this to work all products of the form $1/\epsilon\times\epsilon$
must cancel, otherwise the parts with $A^\epsilon$, defined in
 (\ref{splitintegrals}), would contribute.
We have checked this cancellation explicitly. We have of course also checked
that all divergencies cancel, except those that need to be absorbed in the
new LECs involving the singlet vector current.

The full expression written explicitly in terms of infinite volume and
 finite volume integrals is obtained by expanding the expressions below and
 in Appendix \ref{app:results} using (\ref{eq:mathcalV}) and the corresponding
 expressions for the other integrals. In order to access the finite volume
 corrections any term containing no finite volume integral should be dropped.
 The infinite volume result is obtained by removing all finite volume
 integrals. For the cases presented here the resulting infinite volume
 expressions can be written in the form (\ref{eq:lorentzWard}). Note,
 finally, that all expressions are given in terms of lowest order masses.

\subsection{$\Pi_{\pip_v}^{\mathcal{V}\mu\nu}$ at $p^4$}

\begin{align}
  \label{eq:Pip4}
  \Pi_{0\pip_v}^{\mathcal{V}(4)}(q) =\,&
  4B_{21}^{\mathcal{V}}(m_{u\mathcal{S}}^2,m_{\mathcal{S}d}^2,q)
 - 4 B_{1}^{\mathcal{V}}(m_{u\mathcal{S}}^2,m_{\mathcal{S}d}^2,q)
 + B^{\mathcal{V}}(m_{u\mathcal{S}}^2,m_{\mathcal{S}d}^2,q)\,,
\nonumber\\
  \Pi_{1\pip_v}^{\mathcal{V}(4)}(q) =\,&
 -4B_{22}^{\mathcal{V}}(m_{u\mathcal{S}}^2,m_{\mathcal{S}d}^2,q)
 + A^{\mathcal{V}}(m_{u\mathcal{S}}^2)
 + A^{\mathcal{V}}(m_{d\mathcal{S}}^2)\,,
\nonumber\\
  \Pi_{2\pip_v}^{\mathcal{V}(4)\mu\nu}(q) =\,&
 4 B_{23}^{\mathcal{V}\mu\nu}(m_{u\mathcal{S}}^2,m_{\mathcal{S}d}^2,q) 
- 2 q^\nu B_{2}^{\mathcal{V}\mu}(m_{u\mathcal{S}}^2,m_{\mathcal{S}d}^2,q)
- 2 q^\mu B_{2}^{\mathcal{V}\nu}(m_{u\mathcal{S}}^2,m_{\mathcal{S}d}^2,q)\,.
\end{align}

\subsection{$\Pi_{XY}^{\mathcal{V}\mu\nu}$ at $p^4$}

\begin{align}
  \label{eq:PiXYp4}
  \Pi_{0XY}^{\mathcal{V}(4)}(q) =\,&
 -4B_{21}^{\mathcal{V}}(m_{XY}^2,m_{XY}^2,q)
 + 4 B_{1}^{\mathcal{V}}(m_{XY}^2,m_{XY}^2,q)
 - B^{\mathcal{V}}(m_{XY}^2,m_{XY}^2,q)\,,
\nonumber\\
  \Pi_{1XY}^{\mathcal{V}(4)}(q) =\,&
 4B_{22}^{\mathcal{V}}(m_{XY}^2,m_{XY}^2,q)
 - 2A^{\mathcal{V}}(m_{XY}^2)\,,
\nonumber\\
  \Pi_{2XY}^{\mathcal{V}(4)\mu\nu}(q) =\,&
 -4 B_{23}^{\mathcal{V}\mu\nu}(m_{XY}^2,m_{XY}^2,q)\,.
\end{align}

\subsection{$\left<\bar q \gamma^\mu q\right>^{\mathcal{V}}$ at $p^4$}

\begin{align}
\label{eq:VEVp4}
  \left<\bar q \gamma^\mu q\right>^{\mathcal{V}(4)} =\,& 2A^{\mu\mathcal{V}}(m_{q\mathcal{S}}^2).
\end{align}

\section{Connected versus disconnected}
\label{sec:discon}

In Ref.~\cite{Bijnens:2016ndo} we presented arguments for the ratio of disconnected to connected contributions to vector two-point functions relevant to HVP. The basic observation used was that the singlet vector current does not couple to mesons until $\bO(p^6)$, or $\bO(p^4)$ through the WZW term. In this section we outline how PQ changes the conclusions in that paper.

To discuss the singlet vector current couplings in PQ QCD we need to briefly introduce the supersymmetric formulation of PQ QCD. In this formulation there are three quarks for every single quark in standard QCD. There are two fermionic quarks with different masses, these are the sea and valence quarks. The third quark is a boson with the same mass as the valence quark. Sea quark contributions are associated with closed quark loops. The fermionic and bosonic valence quark closed loop contributions cancel since they contribute with opposite signs. Using correlators formed from valence quarks then leads to PQ QCD.

The singlet vector current in the supersymmetric formulation is
\begin{align}
  V^\mu_s = j^\mu_U + j^\mu_D + j^\mu_S + j^\mu_{\tilde U} + j^\mu_{\tilde D} + j^\mu_{\tilde S} + j^\mu_1 + j^\mu_2 + j^\mu_3,
\end{align}
where $U,D,S$ indicate valence quarks, $\tilde U, \tilde D, \tilde S$ indicate ghost quarks which cancel normal valence quark loops and $1,2,3$ indicate sea quarks. A general feature of two-point functions in the PQ theory is then that
\begin{align}
  \label{eq:PQrel}
  \Pi_{U\tilde U} = - \Pi^D_{UU},\quad \textrm{and} \quad \Pi_{\tilde U} = - \Pi^C_{UU} + \Pi^D_{UU},
\end{align}
where the superscripts, $C$ and $D$, indicate the connected and disconnected part respectively. This follows from the observation that any normal quark loop gives a minus sign whereas bosonic quark loops don't. The connected (disconnected) part of any two-point function contains one (two) valence quark loops which gives the above relations. All other quark loops are in common between the quark and ghost quark currents.

We now turn to the issue of the ratio between disconnected and connected two-point functions. For any two-point function $\Pi_{ab}$ we denote the part which contains only vertices with no coupling to the singlet vector current by $\widetilde\Pi_{ab}$. $\widetilde\Pi_{ab}$ contains, but is not limited to, diagrams which contain vertices only from the $p^2$ and $p^4$ Lagrangians, with the exception of the WZW term. The property that there is no coupling to the singlet vector current gives in the two flavor case
\begin{align}
  \widetilde\Pi_{U(U+D+\tilde U + \tilde D + 1 + 2)} = 0.
\end{align}
Using (\ref{eq:PQrel}) and working in the isospin limit gives
\begin{align}
  \frac{\widetilde\Pi_{U1}}{\widetilde\Pi_{\pip}} = -\frac{1}{2}.
\end{align}
Changing $1\rightarrow D$ gives the unquenched result from \cite{Bijnens:2016ndo}. The PQ theory gives a relation between the connected part with external valence quarks and the disconnected part with one external valence quark and one external sea quark. 

Similarly, the three flavor case in the isospin limit gives the relation
\begin{align}
  \frac{\widetilde\Pi_{U1}}{\widetilde\Pi_{\pip}} =
 -\frac{1}{2} - \frac{\widetilde\Pi_{U3}}{2\widetilde\Pi_{\pip}}.
\end{align}

\section{Disconnected and connected for the strange quark contribution}
\label{disconnectedstrange}

The expressions given in section \ref{sec:analytical} and
Appendix \ref{app:results} and the numerical
results presented below are with lowest order masses. For this reason, low
energy constants related to mass corrections appear in the
two-point functions. In this and the following section we have used
as input for the lowest order masses and decay constant
\begin{align}
  \label{eq:input}
  &m_\pi = 135~\textrm{MeV},\qquad &&m_K = 495~\textrm{MeV},\qquad &&F_\pi = 92.2~\textrm{MeV}\,.
\end{align}
For the LECs we use the values of \cite{Bijnens:2014lea}:
\begin{align}
  L_4^r =\,& 0.3\times 10^{-3},  & L_5^r =\,& 1.0\times 10^{-3}, &
L_6^r =\,& 0.1\times 10^{-3}\ & \mu =\,& 770~\textrm{MeV},
\nonumber\\
  L_8^r =\,& 0.5\times 10^{-3},  &L_9^r =\,& 5.9\times 10^{-3},&
L_{10}^r =\,& -3.8\times 10^{-3},
\end{align}
where $\mu$ is the renormalization scale.

In our earlier work \cite{Bijnens:2016ndo} we estimated the ratio of
disconnected to connected contributions for the two-point functions
with the up and down quark part of the electromagnetic currents. In addition,
we estimated
the size of the contributions from the strange quark electromagnetic current,
$\hPi_S$,
and the mixed strange quark-- up-down quarks, $\hPi_{US}$.
The latter is purely disconnected. We did not estimate the size of the
disconnected contribution for the strange case since in \cite{Bijnens:2016ndo}
we used standard ChPT in the isospin conserving case which did not allow us
to do that. Here we calculated the contributions using
PQChPT so we can now estimate separately the connected and disconnected part.

The arguments for $\widetilde\Pi_{US}=(-1/2)\widetilde\Pi_S$ as given
in \cite{Bijnens:2016ndo} and in section~\ref{sec:discon} remain valid and
 we obtain the same ratios here.

\begin{figure}
\begin{minipage}{0.49\textwidth}
\includegraphics[width=0.99\textwidth]{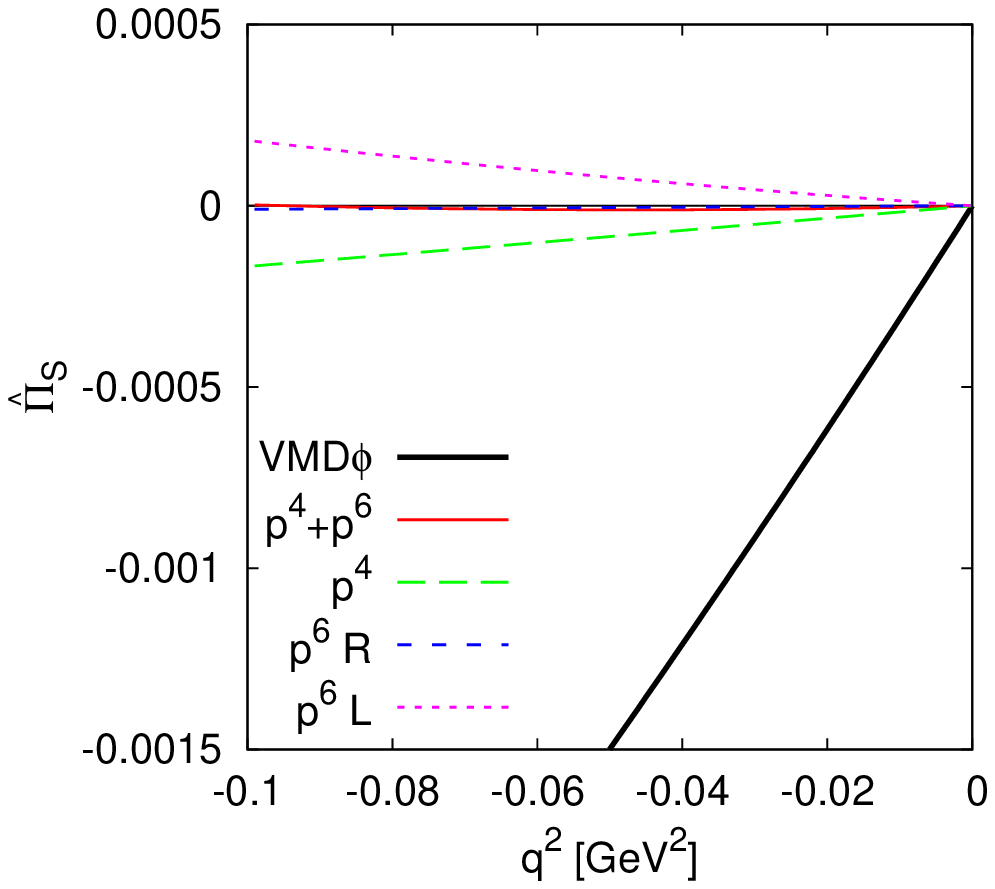}
\centerline{(a)}
\end{minipage}
\begin{minipage}{0.49\textwidth}
\includegraphics[width=0.99\textwidth]{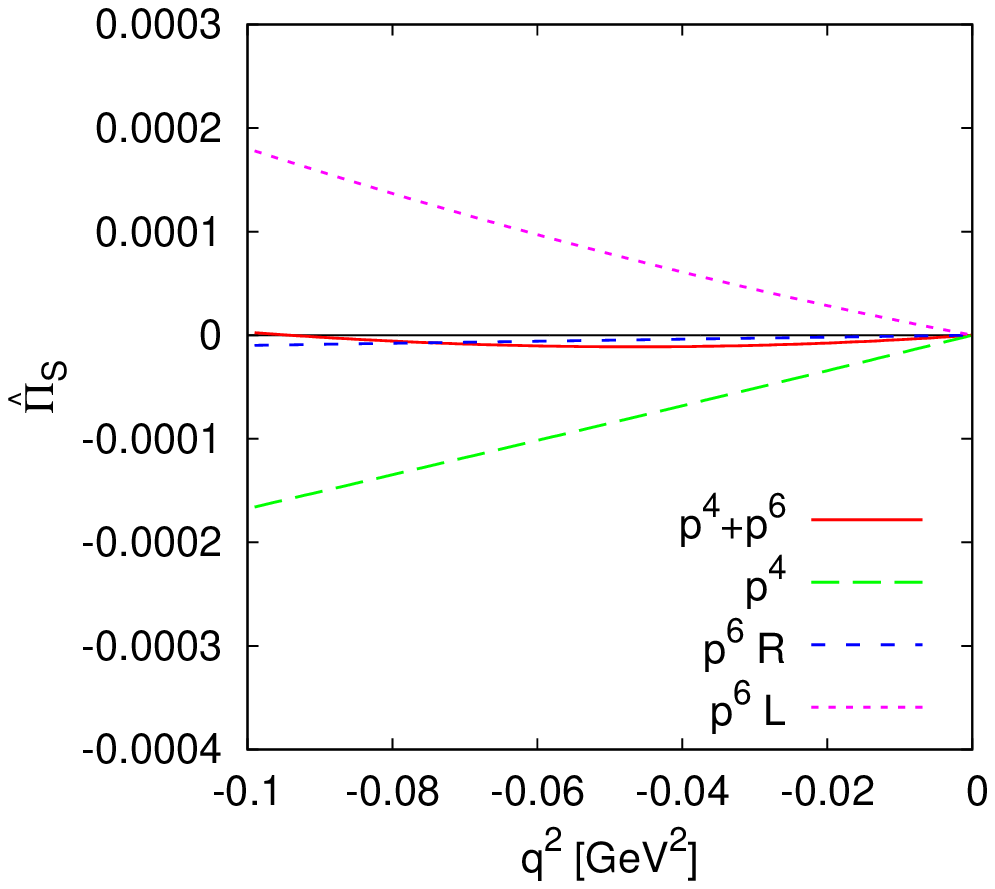}
\centerline{(b)}
\end{minipage}
\caption{(a) The different contributions to $\hPi_S(q^2)$.
The $p^4$ calculation, the pure two-loop part, $p^6 R$, the $p^6$ part
depending on the $p^4$ LECs, $p^6 L$, and the pure LEC contribution
as estimated in \cite{Bijnens:2016ndo} using $\phi$-dominance, $VMD\phi$.
(b) The different loop contributions only, i.e. the $VMD\phi$ contribution 
not included, with the same vertical scale as used in Fig.~\ref{figpis2} but with 
a different range.}
\label{figpis}
\end{figure}

\begin{figure}
\begin{minipage}{0.49\textwidth}
\includegraphics[width=0.99\textwidth]{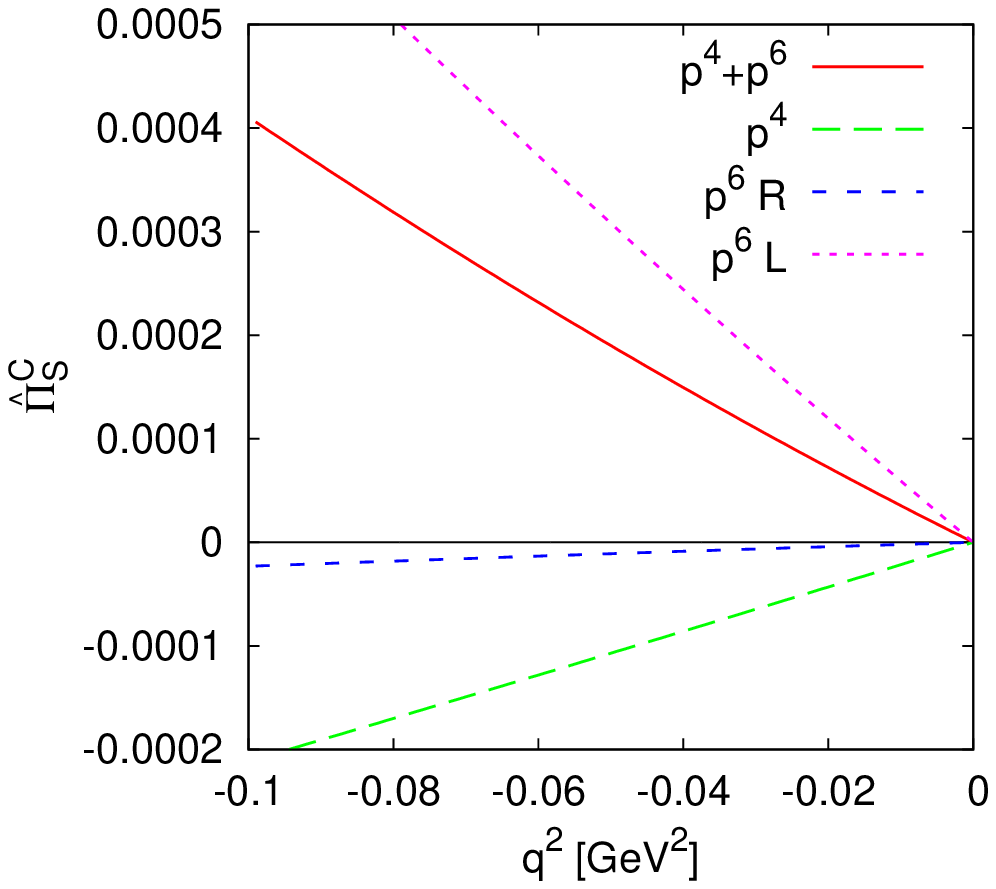}
\centerline{(a)}
\end{minipage}
\begin{minipage}{0.49\textwidth}
\includegraphics[width=0.99\textwidth]{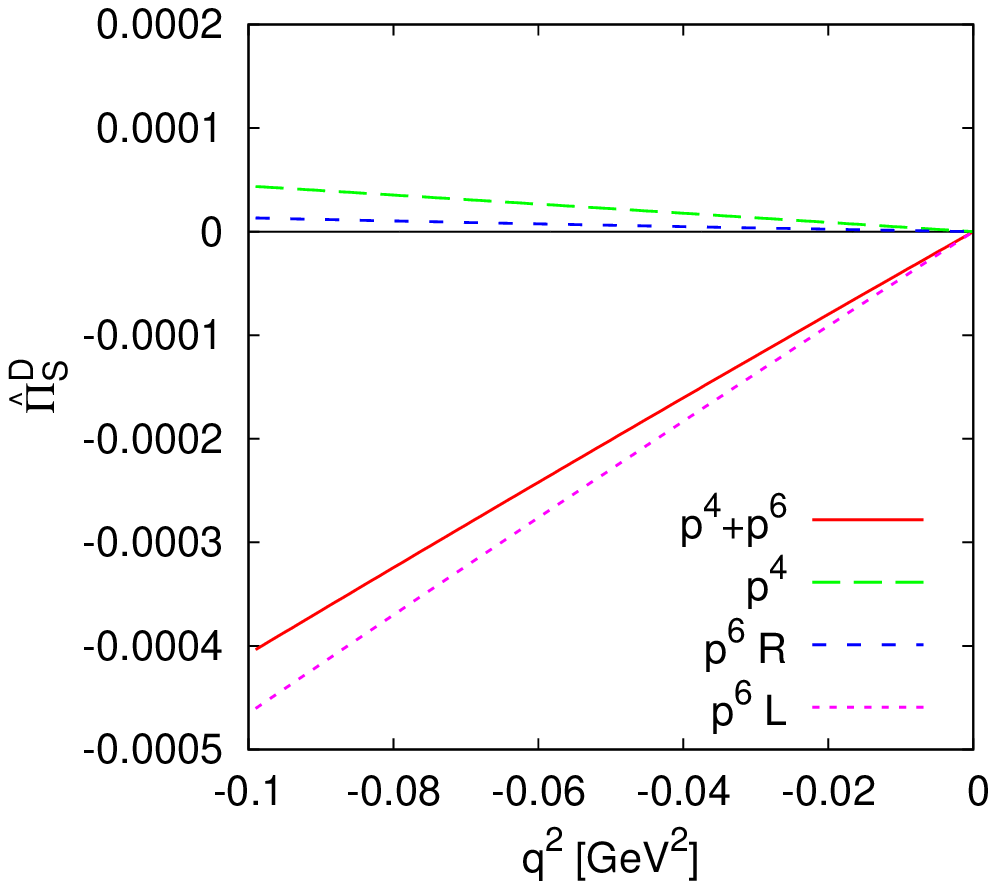}
\centerline{(b)}
\end{minipage}
\caption{(a) The different contributions to the connected part, $\hPi^C_S(q^2)$.
The $p^4$ calculation, the pure two-loop part, $p^6 R$, and the $p^6$ part
depending on the $p^4$ LECs, $p^6 L$. The pure LEC contribution
as estimated by $VMD\phi$ is not shown.
(b) The different contributions to the disconnected part $\hPi^D_S(q^2)$.
The $VMD\phi$ contribution is zero for this case.}
\label{figpis2}
\end{figure}

In Fig.~\ref{figpis}(a) we show the results as obtained
in our earlier work for $\hPi_S(q^2)$ but here in terms of lowest order masses.
It should be remembered that the pure LEC contribution, i.e. tree level
diagrams with no loops, is estimated by $\phi$-meson exchange
and only contributes to $\hPi_S$ and not to $\hPi_{US}$.
For the loop contributions the relation $\hPi_{US}=(-1/2)\hPi_S$
as derived in \cite{Bijnens:2016ndo} holds.
There is a large cancellation between the $p^4$ and $p^6$ contributions
and the final result is very much dominated by the pure LEC contribution
as estimated by $\phi$-exchange.
In Fig.~\ref{figpis}(b) we show the loop contributions with a smaller scale.
For ease of comparison the vertical scale is the same as used in 
Fig.~\ref{figpis2} but with a different range.

In Fig.~\ref{figpis2} the loop contributions for the connected, (a),
and disconnected, (b), parts are shown. It is clear that there is
no simple ratio here as for the up-down case but in all cases the disconnected
contribution is of opposite sign to the connected one and there are
significant cancellations. 

The conclusion here is that the disconnected contribution is of order $-15\%$
of the total strange quark contribution with a sizable error. The error is
both due to the large $p^6$ contribution and the uncertainty on the
$VMD\phi$ estimate. The total strange quark
contribution is by far dominated by the $VMD\phi$ part because even
if individual loop contributions are of order $20\%$, there are large
cancellations making the total strange quark contributions from the loops very
small.

\section{Numerical size of finite volume corrections}
\label{sec:numerics}

In this section we give numerical estimates of the finite volume effects
for vector two-point functions and vacuum expectation values.
In particular we address the questions of convergence of the finite volume
corrections and the effects of using different twist angles for determining
finite volume effects from lattice data. Note that we treat the time direction as infinite.
The numerical input is the same as in section~\ref{disconnectedstrange}
except we have added
\begin{align}
m_\pi L = 4\,.
\end{align}

Our results are for the case with an infinite extent in the time direction.
In realistic lattice calculations the time extent is often twice as large as
the spatial directions and finite volume corrections fall approximately
exponentially with the extent for values considered here.
Our results are thus expected to be a reasonable
approximation to the actually used lattice calculations.
The programs we have used are for a general (Minkowski) time component of $q$
but below we only present numerical results for $q^0=0$.

\subsection{Vector vacuum-expectation-value}

\begin{figure}
\begin{minipage}{0.49\textwidth}
\includegraphics[width=0.99\textwidth]{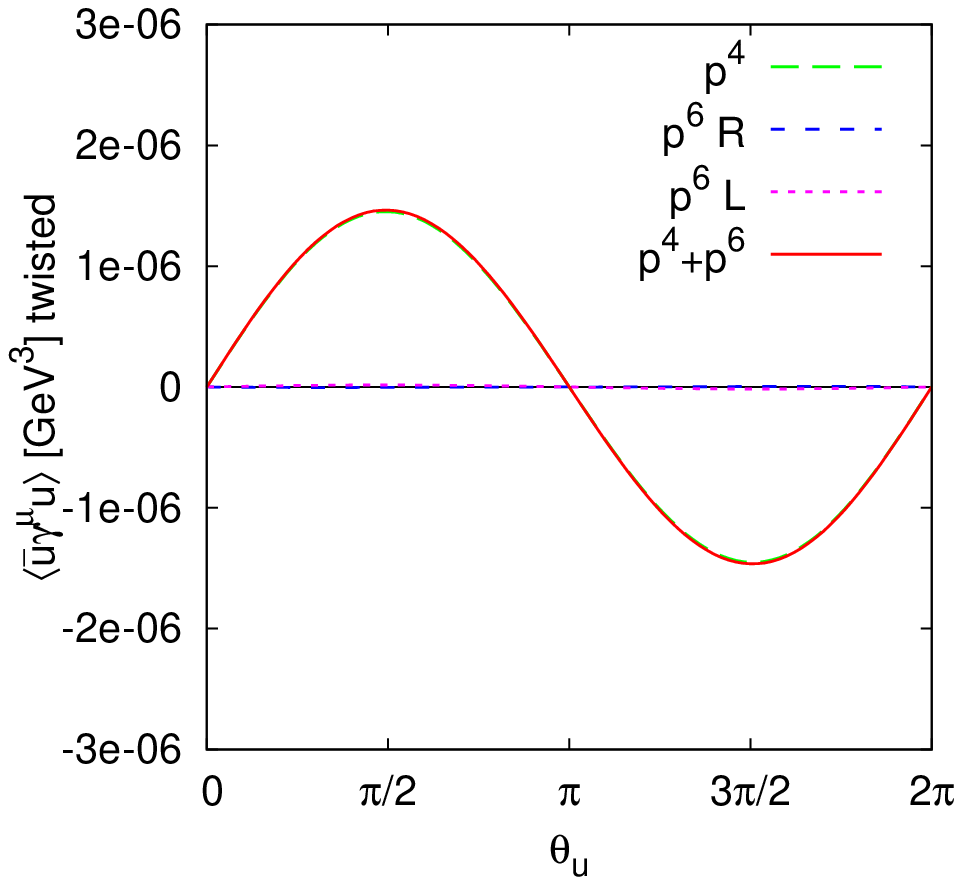}
\centerline{(a)}
\end{minipage}
\begin{minipage}{0.49\textwidth}
\includegraphics[width=0.99\textwidth]{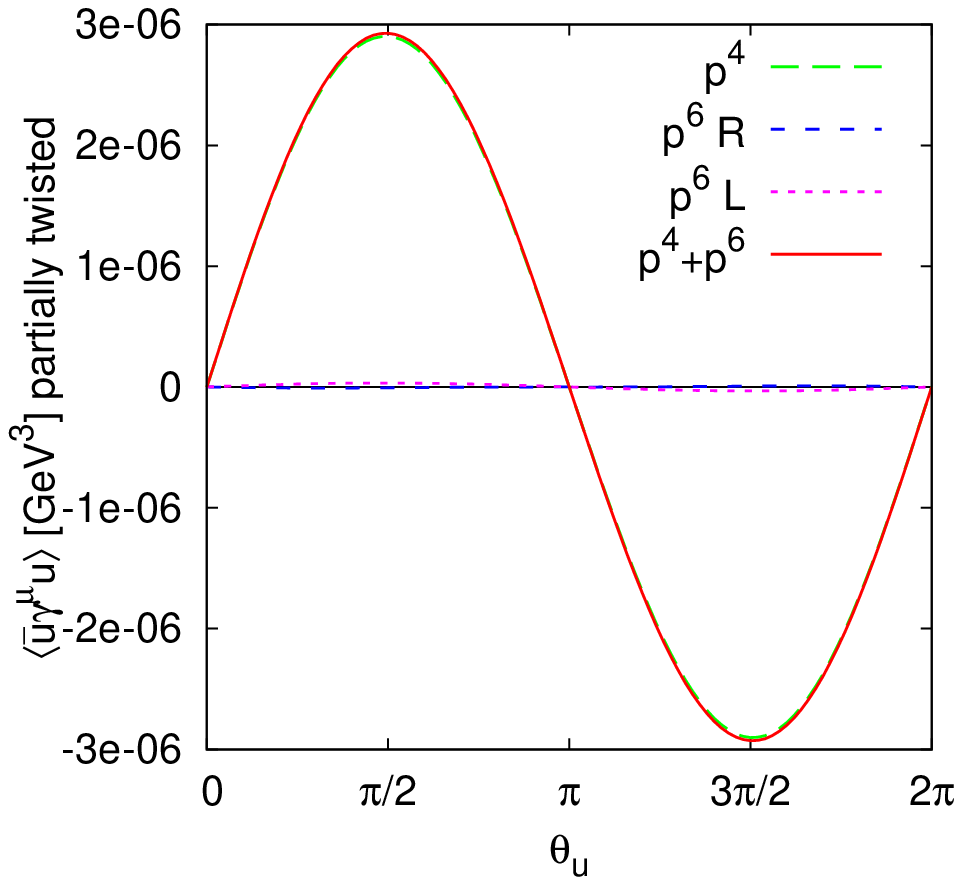}
\centerline{(b)}
\end{minipage}
\caption{(a) The vacuum expectation value $\left<\bar u \gamma^\mu u\right>$
with the up valence and sea quark twisted with $\theta_u=(0,\theta,0,0)$.
(b) Same but only the up valence quark twisted.
In both cases the $x$-component or $\mu=1$ is plotted, the others vanish.
The $p^4$ line is essentially indistinguishable from the $p^4+p^6$ line.}
\label{figvev1}
\end{figure}

As discussed in \cite{Aubin:2013daa,Bijnens:2014yya},
with twisted boundary conditions the vector currents can get a
vacuum expectation value. The one loop result in standard \chpt{} was worked out
in \cite{Bijnens:2014yya}. Here we add the two loop results as well as
partial quenching and twisting. The formulas (\ref{eq:VEVp4}) and (\ref{eq:VEVp6})
are fully general
but we present numerics here for the case where up and down masses are the
same and sea and valence masses equal.
To put the numbers in perspective we can compare with the results
for the scalar vacuum expectation value. The finite volume corrections
here are taken with zero twist using the results of \cite{Bijnens:2006ve}
\begin{align}
\label{vevfiniteV}
\left< \bar u u\right> =\,& -1.2~10^{-2}~\textrm{GeV}^{-3}, &
\left< \bar u u\right>^V(p^4) =\,& -2.4~10^{-5}~\textrm{GeV}^{-3}, 
\nonumber\\
\left< \bar u u\right>^V(p^6R) =\,& 4.5~10^{-7}~\textrm{GeV}^{-3}, &
\left< \bar u u\right>^V(p^6L) =\,& -1.2~10^{-7}~\textrm{GeV}^{-3}. &
\end{align}
In Fig.~\ref{figvev1}(a) we plotted the result for
$\left<\bar u\gamma^\mu u\right>$ for  $\theta_u=(0,\theta,0,0)$
for the fully twisted case, i.e. both the sea and valence up quarks
are twisted. In Fig.~\ref{figvev1}(b) we plot with the same twist
angle but for the partially twisted case, only the up valence quark is twisted.
The finite volume corrections are roughly an order of magnitude smaller
than for the scalar case in (\ref{vevfiniteV}), but the same pattern
is there. The $p^6$ corrections are very small.
The partially twisted case is almost exactly a factor of two larger
than the fully twisted case. The effects are strongly dominated by
the pion loops and for these the difference at $p^4$ is exactly a factor
of two. The vacuum expectation value
$\left<\bar d\gamma^\mu d\right>$ with the up-quark
fully twisted and no twist on the down quark
is almost exactly minus $\left<\bar u\gamma^\mu u\right>$.
Again it is exactly minus for the pion loops only.
For the partially twisted up-quark $\left<\bar d\gamma^\mu d\right>$ 
vanishes since then no active quark has twist.

\subsection{Finite volume corrections for the connected part} 

We now turn to the two-point functions. In the finite volume case
we cannot simply present the combination $\hPi(q^2)$ since the
subtraction at zero is not well defined, after all
$\Pi^{\mu\nu}(q=0)\ne0$. The relevant two-point function to use with
twisted boundary conditions is the connected light part, $\Pi_{\pi^+}$.
In the following we only twist the up-quark. We also put the up and down
masses equal and sea and valence masses the same.

There is essentially no numerical difference between the fully 
twisted (both valence and sea up quark twisted) and partially 
twisted cases. We therefore present only the partially twisted case in the plots.
The Ward identity is fulfilled in both cases but the right hand side
of (\ref{eq:wardSplitPip}) gets the same numerical value in the fully twisted case from
both the up and down vacuum expectation value, and in the partially twisted case
only from the up vacuum expectation value.
\begin{figure}[!t]
\begin{minipage}{0.49\textwidth}
\includegraphics[width=0.99\textwidth]{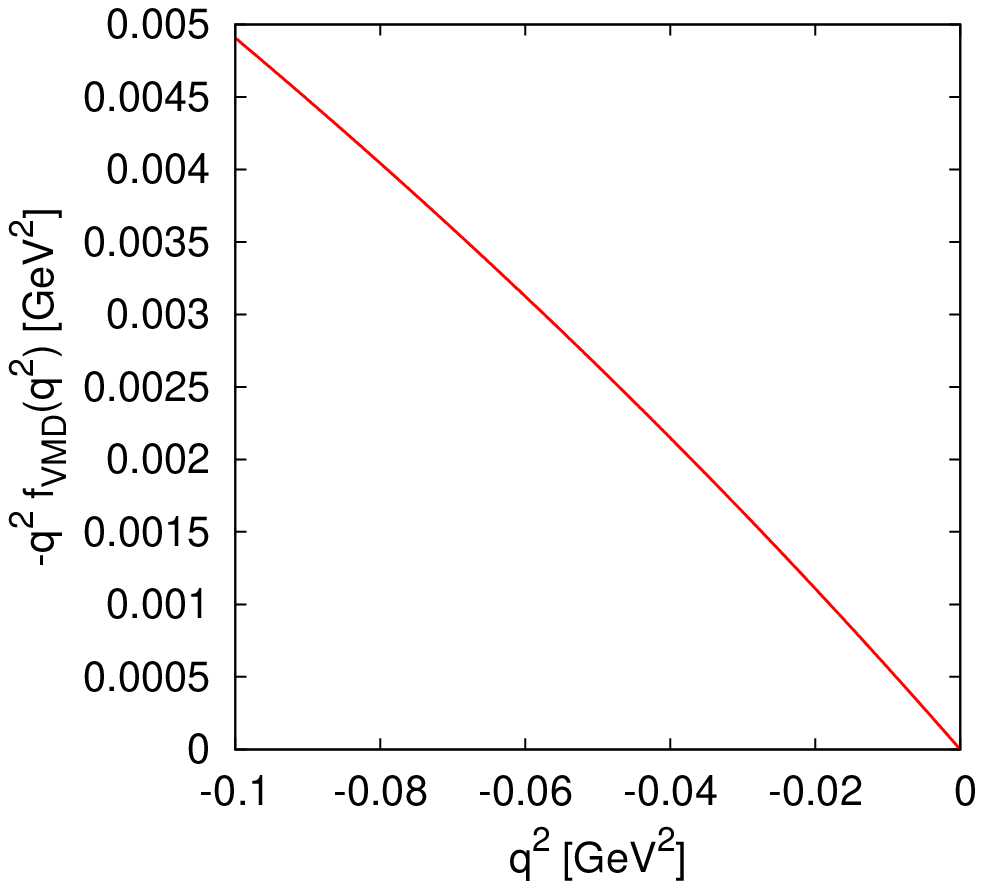}
\centerline{(a)}
\end{minipage}
\begin{minipage}{0.49\textwidth}
\includegraphics[width=0.99\textwidth]{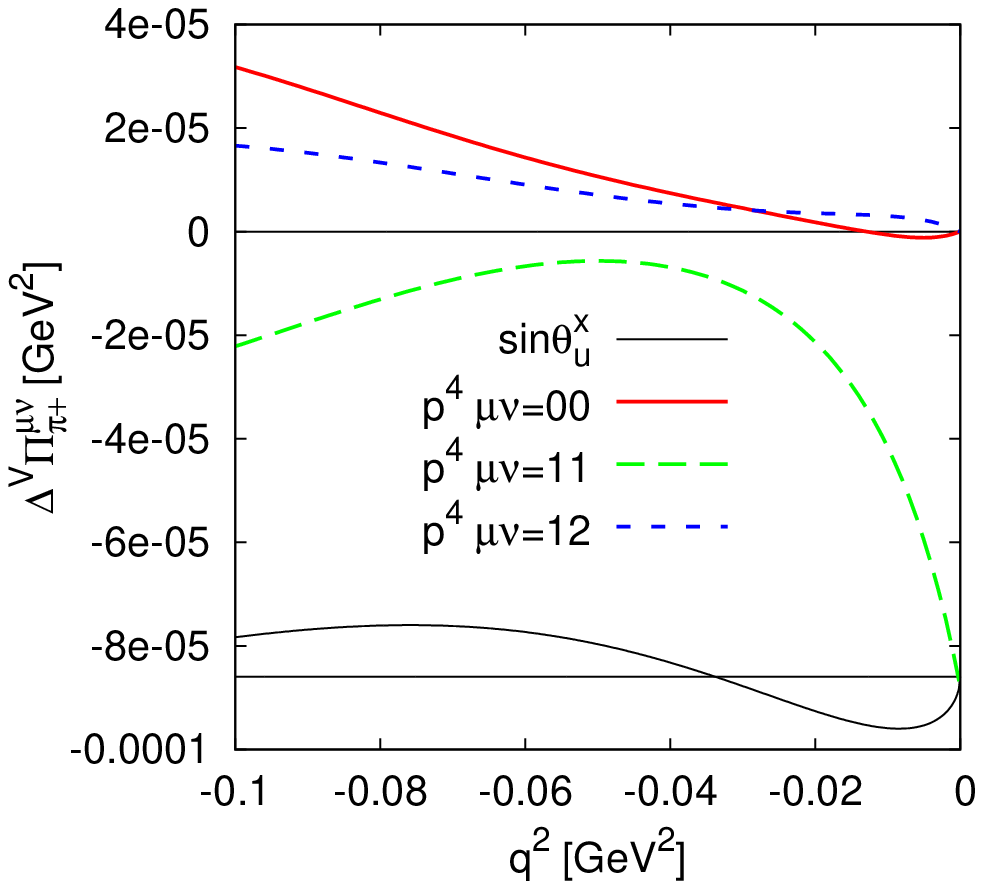}
\centerline{(b)}
\end{minipage}
\caption{(a) $-q^2 f_{\textrm{VMD}}(q^2)$ as a function of $q^2$. This together
with (\ref{VMD1}) and (\ref{VMD2}) can be used to judge the relative size of the
finite volume effects in the following figures.
(b) The finite volume corrections at $p^4$ for the spatially symmetric case.
The lower straight line indicates zero around which $\sin \theta_u^x$ oscillates.}
\label{figVMD}
\end{figure}

In order to show the size of the finite volume
corrections we can compare with the naive VMD estimate.
This corresponds to
\begin{align}
\left.\Pi_{\pi^+}^{\mu\nu}\right|_{\mathrm{VMD}}
= \left(q^\mu q^\nu - q^2 g^{\mu\nu}\right)\frac{4 F_\pi^2}{m_V^2-q^2}
= \left(q^\mu q^\nu - q^2 g^{\mu\nu}\right) f_{\textrm{VMD}}(q^2),
\end{align}
with $m_V=770$~MeV.
When we choose $q=(0,\sqrt{-q^2},0,0)$ we have
\begin{align}
\label{VMD1}
\Pi^{00}= -\Pi^{22} = -\Pi^{33} = -q^2 f_{\textrm{VMD}}(q^2),
\end{align}
and all others zero.
Instead for $q=(0,\sqrt{-q^2/3},\sqrt{-q^2/3},\sqrt{-q^2/3})$
we have that
\begin{align}
\label{VMD2}
\Pi^{00}=\,& -q^2 f_{\textrm{VMD}}(q^2), &
\Pi^{ii} =\,& \frac{2}{3}q^2 f_{\textrm{VMD}}(q^2), &
\left.\Pi^{ij}\right|_{i\ne j} =\,& -\frac{1}{3}q^2f_{\textrm{VMD}}(q^2),
\end{align}
with the others zero.
We have plotted $-q^2 f_{\textrm{VMD}}(q^2)$ in Fig.~\ref{figVMD}(a).

\begin{figure}
\begin{minipage}{0.49\textwidth}
\includegraphics[width=0.99\textwidth]{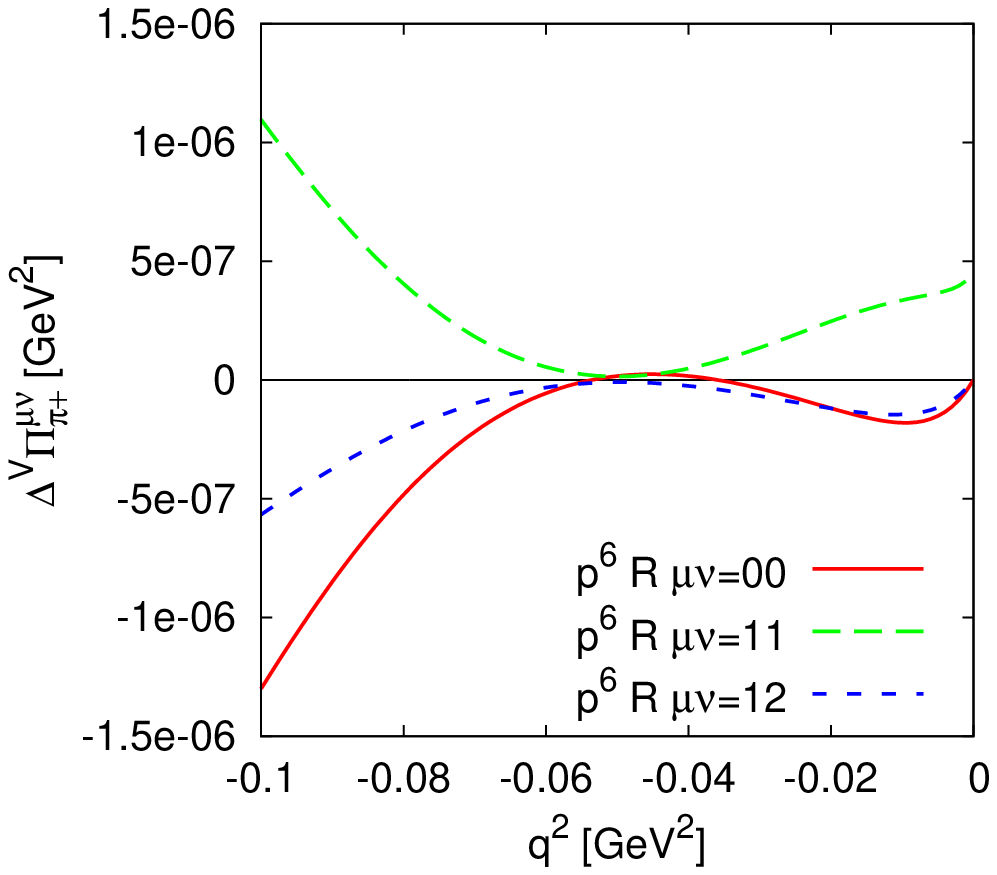}
\centerline{(a)}
\end{minipage}
\begin{minipage}{0.49\textwidth}
\includegraphics[width=0.99\textwidth]{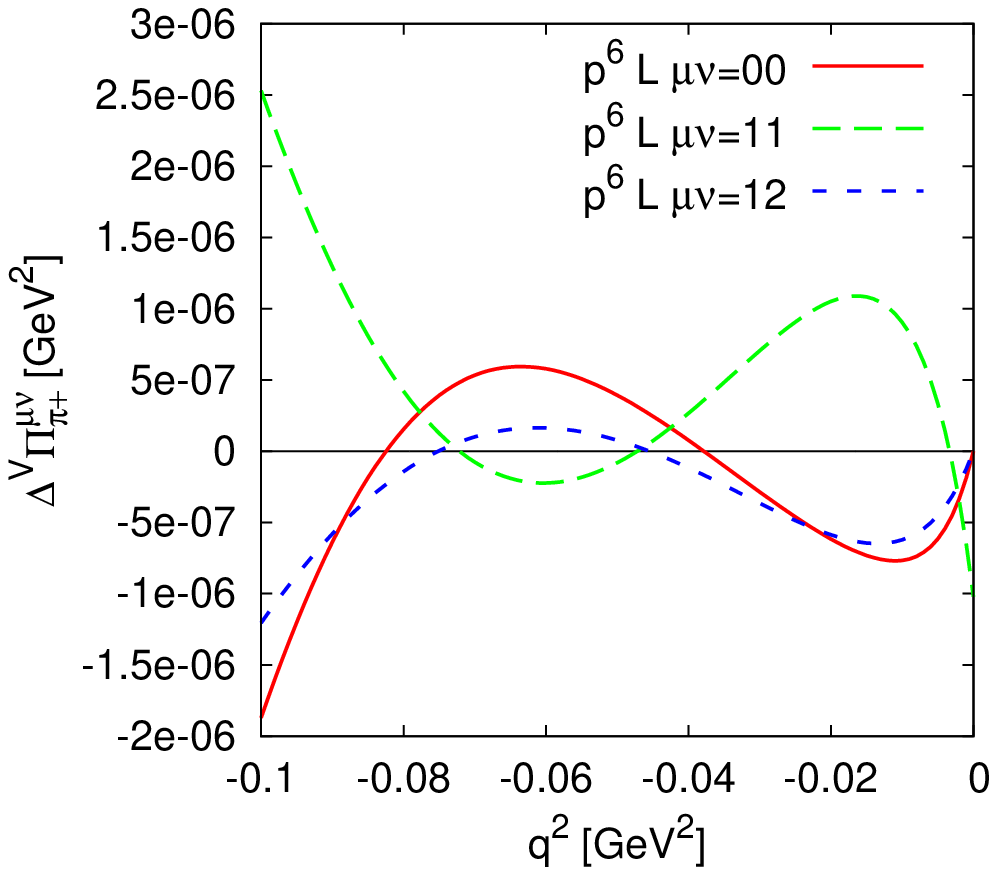}
\centerline{(b)}
\end{minipage}
\caption{The parts of the finite volume corrections at $p^6$ for the spatially
symmetric case(a) $p^6R$ (b) $p^6 L$.}
\label{figxyzp6}
\end{figure}

\begin{figure}
\begin{minipage}{0.49\textwidth}
\includegraphics[width=0.99\textwidth]{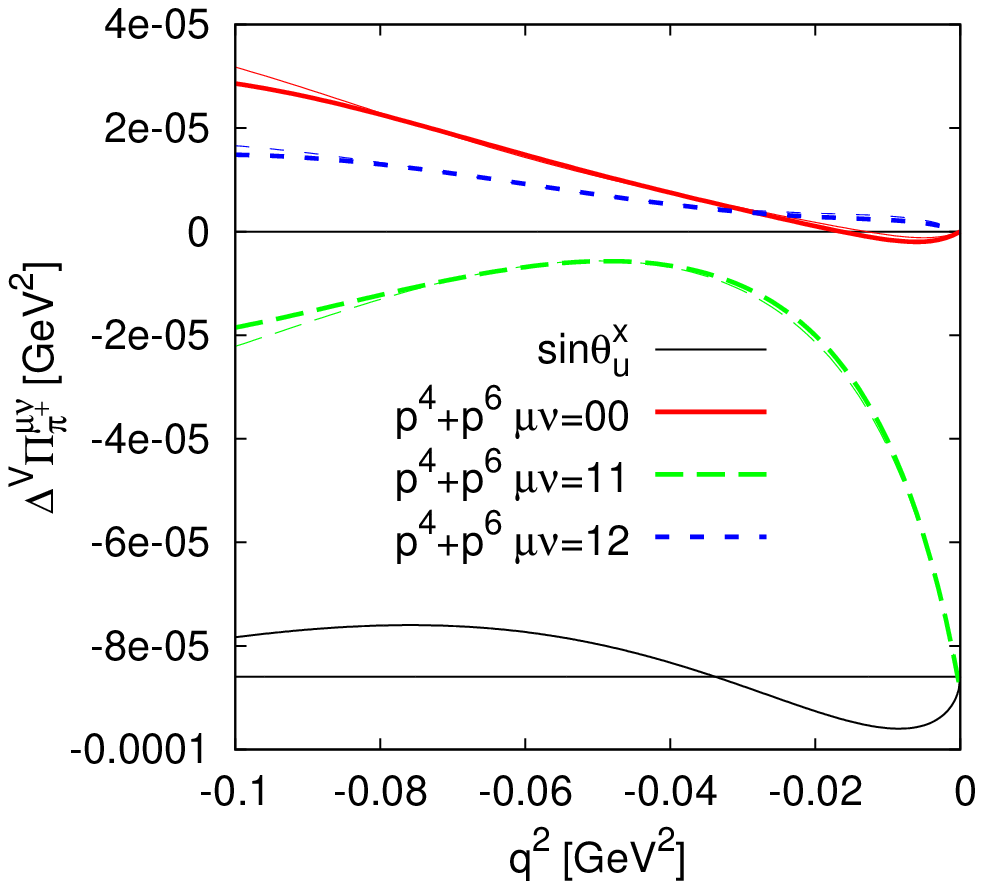}
\centerline{(a)}
\end{minipage}
\begin{minipage}{0.49\textwidth}
\includegraphics[width=0.99\textwidth]{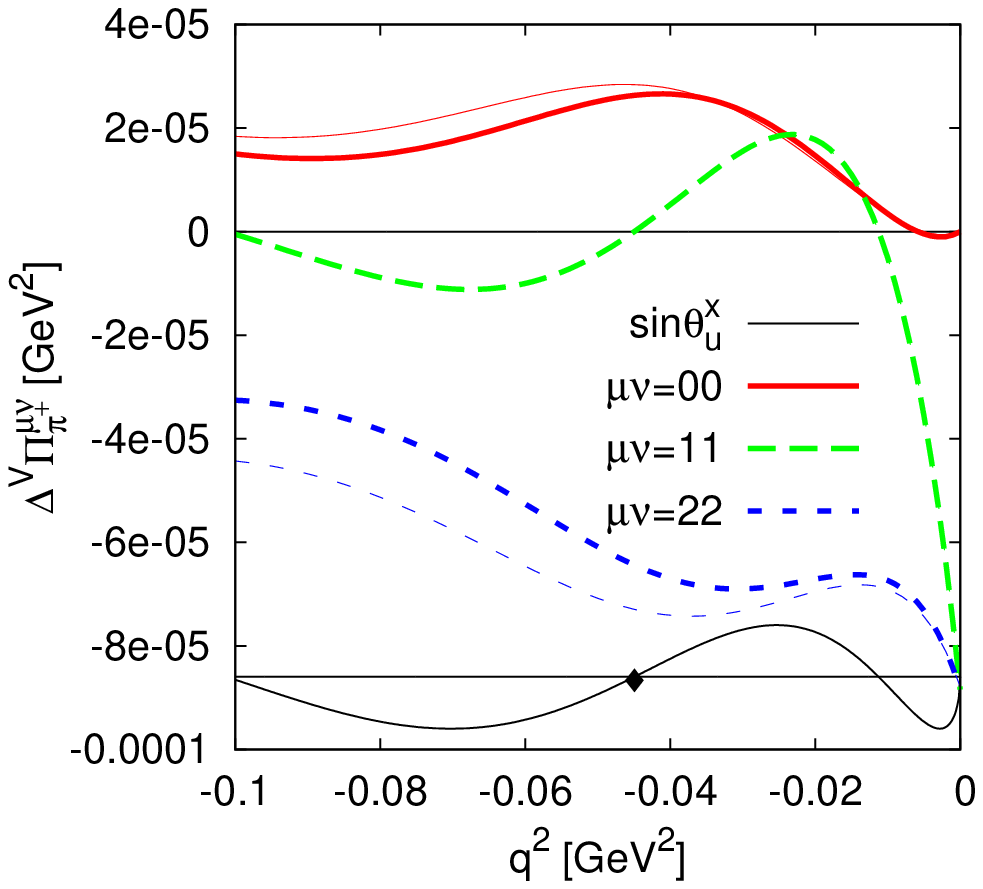}
\centerline{(b)}
\end{minipage}
\caption{The finite volume corrections adding $p^4$ and $p^6$.
The $p^4$ corrections are shown as the thin lines where 
each thin line should be associated with the same pattern and colour
(and closest) thick line. 
The lower straight line indicates zero around which $\sin \theta_u^x$ oscillates.
(a) The spatially
symmetric case (b) Twisting only the $x$-direction.
The diamond indicates a $q^2$ accessible with periodic boundary conditions.}
\label{figp6}
\end{figure}

We can now present the finite volume corrections. First we take the
spatially symmetric twisted case. Here we use $\theta_u=q/L$
with $q=(0,\sqrt{q^2/3},\sqrt{q^2/3},\sqrt{q^2/3})$. The $p^4$
corrections are shown in Fig.~\ref{figVMD}(b). $\Pi^{\mu\nu}(q=0)\ne 0$
is clearly visible. The relative size of the correction compared to the VMD
estimate is in the few \% range (except of course at $q^2=0$ where it becomes
infinite). Note that here we have $\Pi^{11}=\Pi^{22}=\Pi^{33}$,
 $\Pi^{01}=\Pi^{02}=\Pi^{03}=0$ and $\Pi^{12}=\Pi^{13}=\Pi^{23}$.
In \cite{Aubin:2015rzx} they found that lowest order \chpt{} gives a good
description of finite volume effects already at leading order ($p^4$).
If this is the case, then the higher order corrections should turn out to be
small, in contrast to the infinite volume case where they can be significant,
see \cite{Bijnens:2016ndo}. In Fig.~\ref{figxyzp6}
we plot the two parts of the finite volume correction for $\Pi_\pip$
at order $p^6$.
We find that the correction is small, supporting the
conclusion of \cite{Aubin:2015rzx}.
The bottom curves in Fig.~\ref{figVMD}(b) and \ref{figp6}
show $\sin(\theta_u^x)$ allowing to judge the type of twisting effects
expected.

In Fig.~\ref{figp6}(a) we show the full ($p^4+p^6$) finite volume correction
for the spatially symmetric case. The $p^4$ result is included with thin dashed lines for comparison.
Using the same twist angle in all spatial directions
is common in lattice calculations of the HVP.
It gives the possibility to average over several directions reducing the
statistical error. However, the finite volume corrections do depend
on how the twisting is done. We could have chosen to twist only in
the $x$-direction. In that case we have
$\theta_u=q/L$ with $q=(0,\sqrt{q^2},0,0)$ and
$\Pi^{22}=\Pi^{33}$ and all elements with $\mu\ne\nu$ vanish.
The full ($p^4+p^6$) finite volume corrections for this case are shown in Fig.~\ref{figp6}(b).
Again, the $p^4$ results are included with thin dashed lines.

\begin{figure}
\begin{minipage}{0.49\textwidth}
\includegraphics[width=0.99\textwidth]{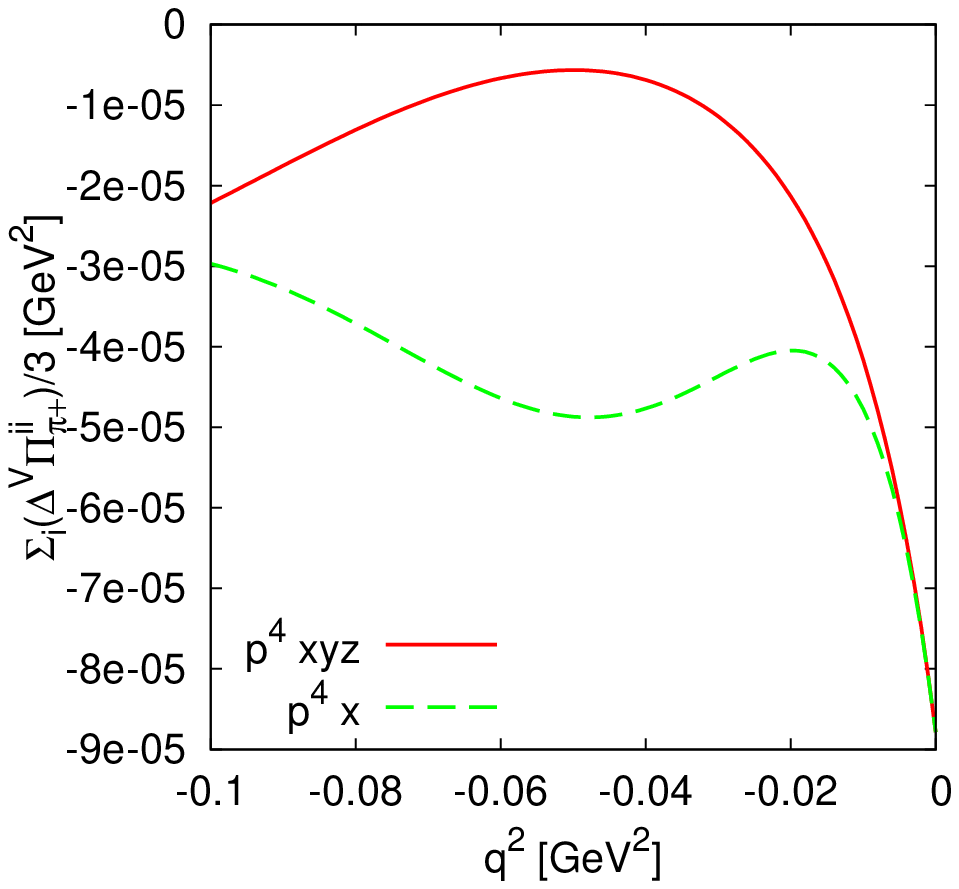}
\centerline{(a)}
\end{minipage}
\begin{minipage}{0.49\textwidth}
\includegraphics[width=0.99\textwidth]{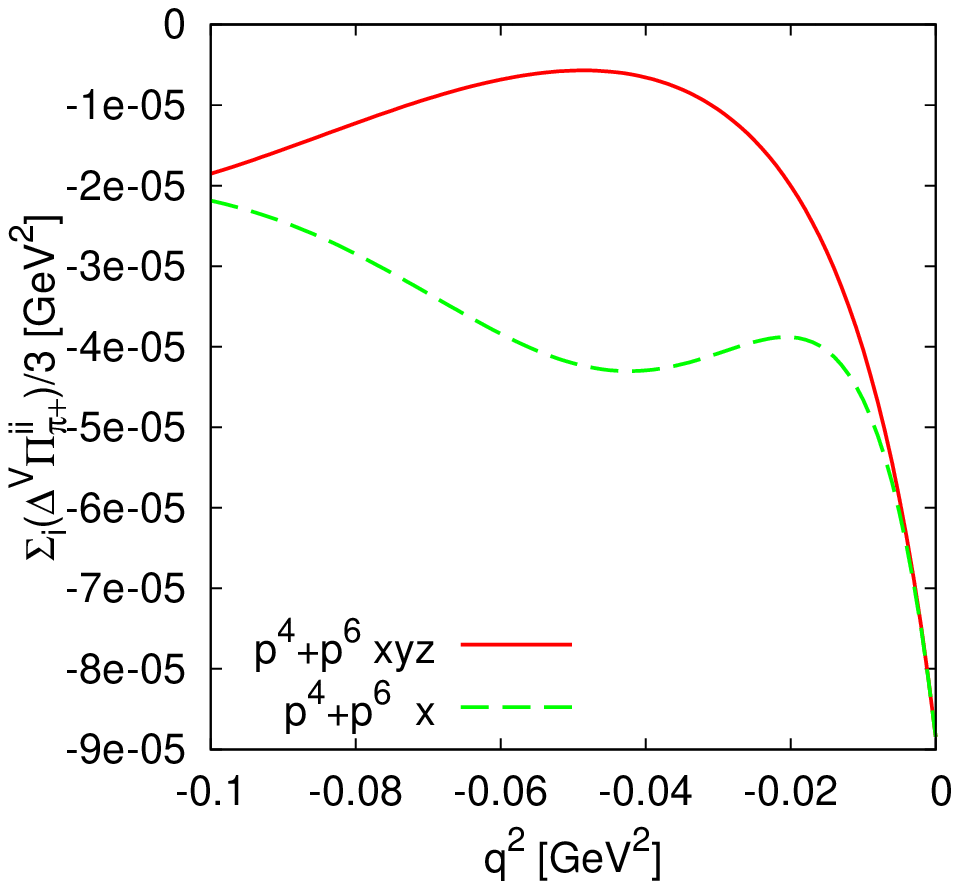}
\centerline{(b)}
\end{minipage}
\caption{The finite volume corrections to the spatial average
as defined in (\ref{defpibar}). $xyz$ is the spatially symmetric twisting and
$x$ twisting only in the $x$-direction.
(a) $p^4$ (b) Sum of $p^4$ and $p^6$.}
\label{figpibar}
\end{figure}

Comparing the two halves of Fig.~\ref{figp6} we see quite different
finite volume corrections. This can be used to test the size of the
finite volume corrections using only lattice data  by using two
different ways of twisting that should reduce to the same $q^2$.
This would also constitute a test of our predictions for the finite volume
corrections. The quantity we will use for this is the average of the 
spatial diagonal components
\begin{align}
\label{defpibar}
\overline\Pi = \frac{1}{3}\sum_{i=1,2,3} \Pi^{ii}\,.
\end{align}
The finite volume corrections to $\overline\Pi$ are shown in
Fig.~\ref{figpibar}. In (a) we show the $p^4$ result and in (b) the sum of
the $p^4$ and $p^6$ results. There is a good convergence and the
difference between spatially symmetric twisting and twisting only
in the $x$-direction is of similar size as the actual correction over
a sizable range of $q^2$. This difference can thus be used to
test the finite volume corrections using the same underlying set of
configurations without having to resort to tricks like reweighting 
\cite{Bussone:2016lty}. That the curves for the two cases coincide
for $q^2=0$ is clear since then the twists vanish fully for both cases.

\subsection{Finite volume corrections for the neutral case and disconnected part}

In the previous subsection we could use (partial-)twisting to obtain any value
of $q^2$ even at finite volume. For the neutral current where the twist on
the quark and anti-quark cancel this is no longer true\footnote{Since we work
with an infinite temporal extension this is not really the case.
We do however give a nonzero value only to the spatial components of $q$ to
give an indication of the size of the finite volume corrections.}.

For $m_\pi L=4$, we only have access to $q^2 =0, -0.045,-0.09$~GeV$^2$
for $|q^2|<0.1$~GeV$^2$. The values for q are respectively
$q=(0,0,0,0)$, $q=(2\pi/L)(0,1,0,0)$ and $q=(2\pi/L)(0,1,1,0)$
(and permutations of $x,y,z$ and changes in signs of components).

The finite volume corrections are dominated by the lightest particle, the pion,
and corrections due to kaons and eta are expected to be small. Our numerical
results confirm this. Finite volume corrections require the
presence of at least one loop in the contributions in ChPT, otherwise there
is no propagating particle to feel the effect of the boundaries. Putting these
two things together, the finite volume corrections are expected to show
the relations between disconnected and connected parts of
$\widetilde\Pi^{\mu\nu}$
in \cite{Bijnens:2016ndo} and Sect.~\ref{sec:discon}
for the three-flavour case exactly and the relations for the two-flavour case
to a very good precision. This is indeed the case for our results.
We thus only quote results for the connected contribution. The disconnected
is $-1/2$ this and the sum of the two $+1/2$ the quoted numbers within the
precision shown in Tab.~\ref{tab:FV}.

The Ward identity $q_\mu \Pi^{\mu\nu}=0$ is also satisfied,
relating a number of the nonzero-components. 
This together with the symmetries determines the nonzero components
not listed. The relations are given in the last column in Tab.~\ref{tab:FV}.
The size of the corrections should
be compared with the results in Fig.~\ref{figVMD}(a) as discussed
in the previous subsection. The value at $q=(0,0,0,0)$ agrees of course with
those shown in the figures for the connected contribution.
The size of the corrections is similar to the case with twist discussed earlier.
\begin{table}
\begin{center}
\newcommand{\m}{\phantom{$-$}}
\begin{tabular}{cccccl}
\hline
$q/(2\pi/L)$ & & $\Pi^{00}_U$ & $\Pi^{11}_U$ & $\Pi^{33}_U$ &  \\
      & & $[10^{-5}~$GeV$^2]$ & $[10^{-5}~$GeV$^2]$ & $[10^{-5}~$GeV$^2]$ &  \\
\hline
\rule{0cm}{13pt}
(0,0,0,0) & $p^4$   & \m0.000 & $-$8.785 & $-$8.785 & $\Pi^{22}_U=\Pi^{33}_U$  \\
          & $p^6~R$ & \m0.000 &  \m0.045 &  \m0.045 &   \\
          & $p^6~L$ & \m0.000 & $-$0.102 & $-$0.102 &   \\
          & sum     & \m0.000 & $-$8.842 & $-$8.842 &   \\
\hline
\rule{0cm}{13pt}
(0,1,0,0) & $p^4$   & \m2.840 &  \m0.000 & $-$7.294 & $\Pi^{22}_U=\Pi^{33}_U$  \\
          & $p^6~R$ &$-$0.091 &  \m0.000 &  \m0.223 &   \\
          & $p^6~L$ &$-$0.117 &  \m0.000 &  \m0.633 &   \\
          & sum     & \m2.632 &  \m0.000 & $-$6.438 &   \\
\hline
\rule{0cm}{13pt}
(0,1,1,0) & $p^4$   & \m3.604 & $-$2.415 & $-$6.442 & $\Pi^{22}_U=\Pi^{11}_U$  \\
          & $p^6~R$ &$-$0.195 &  \m0.128 &  \m0.338 & $\Pi^{12}_U=-\Pi^{11}_U$  \\
          & $p^6~L$ &$-$0.458 &  \m0.376 &  \m1.144 & $\Pi^{21}_U=-\Pi^{11}_U$ \\
          & sum     & \m2.951 & $-$1.911 & $-$4.960 &   \\
\hline 
\end{tabular}
\end{center}
\caption{The finite volume corrections to the components of $\Pi^{\mu\nu}_U$
for the connected part for low lying momenta available with periodic boundary 
conditions. The disconnected part is essentially $-1/2$ the connected part
as discussed in the text. The last columns indicate components which are
related to the explicitly presented ones. Components not mentioned are zero.}
\label{tab:FV}
\end{table}

The use of different twists with the same $q^2$ has been discussed
above as a possible way to test the finite volume corrections without having
to generate a lattice with a different volume. The same can be done here.
Partial twisting does not allow for different values of $q^2$ but it does
give different finite volume corrections since the charged mesons in the loops
will be affected. As an example we have looked at the values for $q=(0,0,0,0)$
but twisting the valence up-quark with a twist
angle $\theta=(0,\pi/2,\pi/2,\pi/2)$. Now the relation between the disconnected
and connected part of about $-1/2$ is no longer valid and our numerics also
show this. The results for the connected and disconnected parts are shown in
Tab.~\ref{tab:FV2}. Given the symmetries of the input we have
here that $\Pi^{00}_U=\Pi^{0i}_U=0$, the spatial diagonal elements are
all the same and the spatial off-diagonal elements are also all the same.
The disconnected part has only small changes compared to the case with no twist
but the connected part changes considerably.
So again, using different twists can be used to check the finite volume
corrections.
\begin{table}
\begin{center}
\newcommand{\m}{\phantom{$-$}}
\begin{tabular}{ccccc}
\hline
\rule{0cm}{13pt}
      & $\Pi^{11}_U$        & $\Pi^{12}_U$   & $\Pi^{11}_U$        & $\Pi^{12}_U$\\
      & $[10^{-5}~$GeV$^2]$ & $[10^{-5}~$GeV$^2]$ & $[10^{-5}~$GeV$^2]$ & $[10^{-5}~$GeV$^2]$ \\
      & \multicolumn{2}{c}{Connected}    & \multicolumn{2}{c}{Disconnected} \\
\hline
\rule{0cm}{13pt}
$p^4$    &  \m0.065 &  \m0.915 &  \m4.392 &    0.000\\
$p^6~R$  & $-$0.012 & $-$0.002 & $-$0.011 &    0.003\\
$p^6~L$  &  \m0.001 &  \m0.014 &  \m0.051 &    0.000\\
sum      &  \m0.054 &  \m0.926 &  \m4.432 &    0.003\\
\hline
\end{tabular}
\caption{The finite volume corrections to the components of $\Pi^{\mu\nu}_U$
for the connected and disconnected part with $q=(0,0,0,0)$ and a partially
twisted up quark with twist angle $\theta=(0,\pi/2,\pi/2,\pi/2)$.
Note the difference with Table~\ref{tab:FV}.} 
\label{tab:FV2}
\end{center}
\end{table}

\section{Conclusion}
\label{sec:conclusion}

In this paper we have calculated the vector one-point and two-point
functions at $p^4$ and $p^6$ using PQ\chpt{} in finite volume with twisted
boundary conditions. We have calculated one connected and one disconnected
two-point function. In PQ\chpt{} this is all that is needed to obtain all
vector two-point functions. The connected two-point function was calculated
by considering a flavor charged current with equal masses. The disconnected
two-point function was calculated using two neutral currents with different
flavors.

Extending the work of \cite{DellaMorte:2010aq} and our work
in \cite{Bijnens:2016ndo} we have used the PQ expressions to give a numerical
estimate of the ratio of disconnected to connected contributions for the
strange quark part of the electromagnetic current. Using VMD for the $\phi$
meson to estimate the pure LEC contribution we obtain a ratio of about $-15\%$.

We have also looked at the effects from finite volume and twisted boundary
conditions. The $p^6$ contributions to the finite volume corrections are
small when compared with the $p^4$ contributions which supports the conclusion
of \cite{Aubin:2015rzx} that $p^4$ describes the observed finite volume
effects. We also point out that the difference between estimates using
different twist angles at the same $q^2$ can be used to estimate
the finite volume corrections using a single lattice volume.

\section*{Acknowledgements}

This work is supported in part by the Swedish Research Council grants
contract numbers 621-2013-4287 and 2015-04089 and by
the European Research Council (ERC) under the European Union's Horizon 2020
research and innovation programme (grant agreement No 668679).
We have made extensive use of \textsc{FORM} \cite{FORM} in this work.

\appendix

\section{Integral notation}
\label{app:integrals}
The loop integrals needed when calculating vector two-point functions are
\begin{align}
  \label{eq:integrals}
  A^{\{\,,\mu,\mu\nu\}}\left([m^2]^n\right) =\,& 
\frac{1}{i}\int_V\frac{d^d k}{(2\pi)^d} \frac{\{1,k^\mu,k^\mu k^\nu\}}{(k^2-m^2)^n}\,,
\nonumber\\
  B^{\{\,,\mu,\mu\nu,\mu\nu\alpha\}}\left([m_1^2]^{n_1},[m_2^2]^{n_2},q\right) =\,& \frac{1}{i}\int_V\frac{d^d k}{(2\pi)^d} \frac{\{1,k^\mu,k^\mu k^\nu,k^\mu k^\nu k^\alpha\}}{(k^2-m_1^2)^{n_1}((q-k)^2-m_2^2)^{n_2}}\,.
\end{align}
When twisted boundary conditions are used the allowed momenta $k$ in $k^2-m^2$ are indicated by the mass, e.g. allowed momenta in $k^2-m_{\pi^+}^2$ are the $\pi^+$ momenta, see \cite{Bijnens:2014yya}. Note that the definitions of the integrals include poles of any order. For poles of order one we use notation where we skip the square brackets and order of the pole, i.e. $A([m^2]^1)\equiv A(m^2)$. 

The integrals above contain both the finite volume and infinite volume contributions. Examplifying with $A$ and suppressing all arguments, we split the integrals according to
\begin{align}
\label{splitintegrals}
  A =\,& \frac{C_{A}}{\bar\epsilon} + A^{\mathcal V} + \epsilon A^{\epsilon} + \mathcal{O}(\epsilon^2),\nonumber\\
  \frac{1}{\bar\epsilon} =\,& \frac{1}{\epsilon} + \ln(4\pi) + 1 - \gamma.
\end{align}
The constant $C_{A}$ is the residue of the $1/\bar\epsilon$ pole and differs from integral to integral. We renormalize our expressions using the \chpt{} version of $\overline{MS}$ where parts proportional to $1/\bar\epsilon$ cancel. $A^{\mathcal V}$ then contains the part of the infinite volume integral which remains after renormalization plus the finite volume correction. We express this as
\begin{align}
  A^{\mathcal V} = \bar A + A^V,
\end{align}
where $\bar A$ is the infinite volume part and $A^V$ is the finite volume correction. 

The infinite volume part of the integrals, including the residues of the poles, can be found from \cite{Bijnens:2002hp} using that the higher pole integrals can be obtained by derivatives with respect to the masses. Methods for evaluating the finite volume correction, as well as expressions for some of the integrals, can be found in \cite{Sachrajda:2004mi,Bijnens:2013doa,Bijnens:2014yya}. In \cite{Bijnens:2014yya} we gave explicit expressions, in terms of Jacobi theta functions, for the finite volume corrections to all of the integrals except for $B^{\mu\nu\alpha}\left([m_1^2]^{n_1},[m_2^2]^{n_2},q\right)$. The expression for the finite volume correction to $B^{\mu\nu\alpha}\left([m_1^2]^{n_1},[m_2^2]^{n_2},q\right)$ is 
\begin{align}
  B^{V\mu\nu\alpha}&\left([m_1^2]^{n_1},[m_2^2]^{n_2},q\right) = 
  \frac{\Gamma(n_1+n_2)}{\Gamma(n_1)\Gamma(n_2)}\int dx (1-x)^{n_1-1}x^{n_2-1}\times\Big(\nonumber\\
  & A^{V\mu\nu\alpha}([\tilde m^2]^{n_1+n_2})
  + x(\delta_\rho^\mu\delta_\sigma^\nu q^\alpha + \delta_\rho^\mu q^\nu\delta_\sigma^\alpha + q^\mu\delta_\rho^\nu\delta_\sigma^\mu)A^{V\rho\sigma}([\tilde m^2]^{n_1+n_2})\nonumber\\
  +&x^2(\delta_\rho^\mu q^\nu q^\alpha + q^\mu \delta_\rho^\nu q^\alpha + q^\mu q^\nu\delta_\rho^\alpha)A^{V\rho}([\tilde m^2]^{n_1+n_2})
  +x^3q^\mu q^\nu q^\alpha A^{V}([\tilde m^2]^{n_1+n_2})\Big)\,,
\end{align}
where 
\begin{align}
  \tilde m^2 = (1-x)m_1^2 + x m_2^2 - x(1-x)q^2,
\end{align}
and the integrals on the right hand side should be evaluated with the
twist angle
\begin{align}
  \vec{\tilde \theta} = \vec\theta -x\vec q L.
\end{align}

In the actual results we have split the integrals as
\begin{align}
  B^{\mu\nu\alpha}& = q^\mu q^\nu q^\alpha B_{31} + (g^{\mu\nu}q^\alpha + g^{\mu\alpha}q^\nu + g^{\nu\alpha}q^\mu) B_{32} + B_{33}^{\mu\nu\alpha},\nonumber\\
  B^{\mu\nu}& = q^\mu q^\nu B_{21} + g^{\mu\nu} B_{22} + B_{23}^{\mu\nu},\nonumber\\
  B^{\mu}& = q^\mu B_{1} + B_{2}^{\mu},\nonumber\\
  A^{\mu\nu}& = g^{\mu\nu} A_{22} + A_{23}^{\mu\nu},
\end{align}
where all arguments are suppressed.

The diagonal integral introduced in (\ref{eq:diagProp}) can in principle be
split up using the residue notation of \cite{Aubin:2003mg} so that all
integrals are of the form (\ref{eq:integrals}). This corresponds to partial
fractioning $\mathcal{D}_{ab}$.
This leads to longer and more difficult to read expressions and we keep the
diagonal propagator intact using notation such as
\begin{align}
  A(\mathcal{D}_{ab}) = \frac{1}{i}\int_V\frac{d^d k}{(2\pi)^d}\left(- \frac{1}{3}\frac{(p^2-m_1^2)(p^2-m_2^2)(p^2-m_3^2)}{(p^2-m_a^2)(p^2-m_b^2)(p^2-m_{\pi^0}^2)(p^2-m_{\eta}^2)}\right).
\end{align}
The residue notation is used in the numerical implementation of our results.

\section{Analytical results}
\label{app:results}

In this appendix we present the analytical expressions for vector two-point
functions and one-point functions at $p^6$ in PQ\chpt{} in finite volume. In
 the case of $\Pi_{\pi^+_v}$ the expressions also contain effects from partially
 twisted boundary conditions. Additionally the expressions contain both 
the infinite volume part and the finite volume correction, see
 section \ref{sec:analytical}, where the $p^4$ expressions are presented.
 Note that all expressions are given using lowest order masses. 
All expressions given will be implemented numerically
 in \textsc{CHIRON} \cite{Bijnens:2014gsa,chiron}.

\subsection{$\Pi_{\pi^+_v}^{\mathcal{V}\mu\nu}$ at $p^6$}
The results presented below are for the various components
 of $\Pi_{\pi^+_v}^{\mathcal{V}(6)}$. These are the infinite and finite
 volume $p^6$ expressions for $\Pi_{\pi^+_v}$. The results are the partially
 twisted ones. Note, however, that the expressions are for the case when
 the two valence quark masses are set equal. The case where the valence
 quark masses differ is much longer and will not be given here.

\begin{align}
  \label{eq:Pi0p6}
  F_0^2&\Pi_{0\pi^+_v}^{\mathcal{V}(6)} = \nonumber
\\
  &+ A^{\mathcal{V}}(m_{x\mathcal{S}}^2)  \Big(
          - 4L^{10}_{r}
          - 4L^9_{r}
          - \frac{1}{2}B^{\mathcal{V}}(m_{x\mathcal{S}^{\prime}}^2,m_{\mathcal{S}^{\prime}y}^2,q)
           + B_1^{\mathcal{V}}(m_{x\mathcal{S}^{\prime}}^2,m_{\mathcal{S}^{\prime}y}^2,q)
          \Big)
\nonumber\\
&       + A^{\mathcal{V}}(m_{y\mathcal{S}}^2)  \Big(
          - 4L^{10}_{r}
          + 4L^9_{r}
          + 4B_{21}^{\mathcal{V}}(m_{x\mathcal{S}^{\prime}}^2,m_{\mathcal{S}^{\prime}y}^2,q)
          + \frac{7}{6}B^{\mathcal{V}}(m_{x\mathcal{S}^{\prime}}^2,m_{\mathcal{S}^{\prime}y}^2,q)
\nonumber\\ &
\hspace*{1cm}          - \frac{13}{3}B_1^{\mathcal{V}}(m_{x\mathcal{S}^{\prime}}^2,m_{\mathcal{S}^{\prime}y}^2,q)
          \Big)
\nonumber\\
&       + \Big(A^{\mathcal{V}}(m_{xy}^2) - A^{\mathcal{V}}(m_{xx}^2)\Big) \Big(
          - 8B_{21}^{\mathcal{V}}([m_{xx}^2]^2,m_{xy}^2,q)m_{xx}^2
          + \frac{8}{3}B^{\mathcal{V}}(m_{xy}^2,m_{xx}^2,q)
\nonumber\\
&\hspace*{1cm}          - 2B^{\mathcal{V}}([m_{xx}^2]^2,m_{xy}^2,q)m_{xx}^2
          - \frac{16}{3}B_1^{\mathcal{V}}(m_{xy}^2,m_{xx}^2,q)
          + 8B_1^{\mathcal{V}}([m_{xx}^2]^2,m_{xy}^2,q)m_{xx}^2
          \Big)
\nonumber\\
&       + A^{\mathcal{V}}(\mathcal{D}_{x\mathcal{S}})  \Big(
          + 4B_{21}^{\mathcal{V}}([m_{x\mathcal{S}}^2]^2,m_{\mathcal{S}y}^2,q)m_{x\mathcal{S}}^2
          + 4B_{21}^{\mathcal{V}}([m_{\mathcal{S}y}^2]^2,m_{x\mathcal{S}}^2,q)m_{x\mathcal{S}}^2
\nonumber\\
&\hspace*{1cm}          + B^{\mathcal{V}}([m_{x\mathcal{S}}^2]^2,m_{\mathcal{S}y}^2,q)m_{x\mathcal{S}}^2
          + B^{\mathcal{V}}([m_{\mathcal{S}y}^2]^2,m_{x\mathcal{S}}^2,q)m_{x\mathcal{S}}^2
          - 4B_1^{\mathcal{V}}([m_{x\mathcal{S}}^2]^2,m_{\mathcal{S}y}^2,q)m_{x\mathcal{S}}^2
\nonumber\\ &
\hspace*{1cm}
          - 4B_1^{\mathcal{V}}([m_{\mathcal{S}y}^2]^2,m_{x\mathcal{S}}^2,q)m_{x\mathcal{S}}^2
          \Big)
\nonumber\\
&       + A^{\mathcal{V}\rho}(m_{x\mathcal{S}}^2)  \Big(
          - 4B_{31}^\mathcal{V}([m_{x\mathcal{S}^{\prime}}^2]^2,m_{\mathcal{S}^{\prime}y}^2,q)q_{\rho}
          + 4B_{21}^{\mathcal{V}}([m_{x\mathcal{S}^{\prime}}^2]^2,m_{\mathcal{S}^{\prime}y}^2,q)q_{\rho}
\nonumber\\ &
\hspace*{1cm}
          - B_1^{\mathcal{V}}([m_{x\mathcal{S}^{\prime}}^2]^2,m_{\mathcal{S}^{\prime}y}^2,q)q_{\rho}
           -B_{2\rho}^{\mathcal{V}}([m_{x\mathcal{S}^{\prime}}^2]^2,m_{\mathcal{S}^{\prime}y}^2,q)
          \Big)
\nonumber\\
&       + A^{\mathcal{V}\rho}(m_{y\mathcal{S}}^2)  \Big(
          + 4B_{31}^\mathcal{V}([m_{\mathcal{S}^{\prime}y}^2]^2,m_{x\mathcal{S}^{\prime}}^2,q)q_{\rho}
          - 4B_{21}^{\mathcal{V}}([m_{\mathcal{S}^{\prime}y}^2]^2,m_{x\mathcal{S}^{\prime}}^2,q)q_{\rho}
\nonumber\\
&\hspace*{1cm}          + B_1^{\mathcal{V}}([m_{\mathcal{S}^{\prime}y}^2]^2,m_{x\mathcal{S}^{\prime}}^2,q)q_{\rho}
           +B_{2\rho}^{\mathcal{V}}([m_{\mathcal{S}^{\prime}y}^2]^2,m_{x\mathcal{S}^{\prime}}^2,q)
          \Big)
\nonumber\\
&       - A^{\mathcal{V}\rho}(m_{\mathcal{S}^{\prime}\mathcal{S}}^2)  \Big(
         + 4B_{31}^\mathcal{V}([m_{x\mathcal{S}}^2]^2,m_{\mathcal{S}y}^2,q)q_{\rho}
          - 4B_{31}^\mathcal{V}([m_{\mathcal{S}y}^2]^2,m_{x\mathcal{S}}^2,q)q_{\rho}
\nonumber\\&\hspace*{1cm}         + 4B_{21}^{\mathcal{V}}([m_{x\mathcal{S}}^2]^2,m_{\mathcal{S}y}^2,q)q_{\rho}
          - 4B_{21}^{\mathcal{V}}([m_{\mathcal{S}y}^2]^2,m_{x\mathcal{S}}^2,q)q_{\rho}
         + B_1^{\mathcal{V}}([m_{x\mathcal{S}}^2]^2,m_{\mathcal{S}y}^2,q)q_{\rho}
\nonumber\\& \hspace*{1cm}
          - B_1^{\mathcal{V}}([m_{\mathcal{S}y}^2]^2,m_{x\mathcal{S}}^2,q)q_{\rho}
         + B_{2\rho}^{\mathcal{V}}([m_{x\mathcal{S}}^2]^2,m_{\mathcal{S}y}^2,q)
          - B_{2\rho}^{\mathcal{V}}([m_{\mathcal{S}y}^2]^2,m_{x\mathcal{S}}^2,q)
          \Big)
\nonumber\\
&       + q^2 B^{\mathcal{V}}(m_{x\mathcal{S}}^2,m_{\mathcal{S}y}^2,q)  \Big(
           +2B_{21}^{\mathcal{V}}(m_{x\mathcal{S}^{\prime}}^2,m_{\mathcal{S}^{\prime}y}^2,q)
            - B_1^{\mathcal{V}}(m_{x\mathcal{S}^{\prime}}^2,m_{\mathcal{S}^{\prime}y}^2,q)
            + \frac{1}{2}B^{\mathcal{V}}(m_{x\mathcal{S}^{\prime}}^2,m_{\mathcal{S}^{\prime}y}^2,q)
\nonumber\\& \hspace*{1cm}
            + 4L^9_{r}\Big)
\nonumber \\
&       + \Big(B^{\mathcal{V}}([m_{x\mathcal{S}}^2]^2,m_{\mathcal{S}y}^2,q) + B^{\mathcal{V}}([m_{y\mathcal{S}}^2]^2,m_{x\mathcal{S}}^2,q)\Big)  \Big(
          + 8m_{x\mathcal{S}}^4\Big(2L^8_r-L^5_r\Big)
\nonumber \\& \hspace*{1cm}
           + 8m_{\mathcal{S}^{\prime}\mathcal{S}^{\prime}}^2m_{x\mathcal{S}}^2\Big(2L^6_r-L^4_r\Big)
        \Big)
\nonumber \\
&       -\frac{4}{3} B^{\mathcal{V}}(m_{x\mathcal{S}^{\prime}}^2,m_{\mathcal{S}^{\prime}y}^2,q)  \Big(
         + B_{21}^{\mathcal{V}}(m_{x\mathcal{S}}^2,m_{\mathcal{S}y}^2,q)q^2
          + B_{22}^{\mathcal{V}}(m_{x\mathcal{S}}^2,m_{\mathcal{S}y}^2,q)
          + B_1^{\mathcal{V}}(m_{x\mathcal{S}}^2,m_{\mathcal{S}y}^2,q)q^2\Big)
\nonumber \\
&       + B^{\mathcal{V}}(m_{xy}^2,m_{xx}^2,q)^2  \Big(
          - \frac{5}{3}q^2
          + 2m_{xx}^2
          \Big)
\nonumber \\&
       + B^{\mathcal{V}}(m_{xy}^2,m_{xx}^2,q)  \Big(
          + \frac{20}{3}B_1^{\mathcal{V}}(m_{xy}^2,m_{xx}^2,q)q^2
          - 8B_1^{\mathcal{V}}(m_{xy}^2,m_{xx}^2,q)m_{xx}^2\Big)
\nonumber\\
&       + q^2B_1^{\mathcal{V}}(m_{x\mathcal{S}}^2,m_{\mathcal{S}y}^2,q)  \Big(
         - 2B_{21}^{\mathcal{V}}(m_{x\mathcal{S}^{\prime}}^2,m_{\mathcal{S}^{\prime}y}^2,q)
          + \frac{5}{3}B_1^{\mathcal{V}}(m_{x\mathcal{S}^{\prime}}^2,m_{\mathcal{S}^{\prime}y}^2,q)
          - 8L^9_{r}
          \Big)
\nonumber\\
&       - \Big(B_1^{\mathcal{V}}([m_{x\mathcal{S}}^2]^2,m_{\mathcal{S}y}^2,q)+B_1^{\mathcal{V}}([m_{y\mathcal{S}}^2]^2,m_{x\mathcal{S}}^2,q)\Big)  \Big(
          + 32m_{x\mathcal{S}}^4\Big(2L^8_r-L^5_r\Big)
\nonumber \\& \hspace*{1cm}
           +  32m_{\mathcal{S}^{\prime}\mathcal{S}^{\prime}}^2m_{x\mathcal{S}}^2\Big(2L^6_r-L^4_r\Big)\Big)
\nonumber\\
&       + B_1^{\mathcal{V}}(m_{x\mathcal{S}^{\prime}}^2,m_{\mathcal{S}^{\prime}y}^2,q)  \Big(
          + \frac{14}{3}B_{21}^{\mathcal{V}}(m_{x\mathcal{S}}^2,m_{\mathcal{S}y}^2,q)q^2
          + \frac{20}{3}B_{22}^{\mathcal{V}}(m_{x\mathcal{S}}^2,m_{\mathcal{S}y}^2,q)\Big)
\nonumber\\
&       + B_1^{\mathcal{V}}(m_{xy}^2,m_{xx}^2,q)^2  \Big(
          - \frac{20}{3}q^2
          + 8m_{xx}^2
\Big)
\nonumber\\
&       -4 B_{21}^{\mathcal{V}}(m_{x\mathcal{S}}^2,m_{\mathcal{S}y}^2,q) \Big(
          +q^2B_{21}^{\mathcal{V}}(m_{x\mathcal{S}^{\prime}}^2,m_{\mathcal{S}^{\prime}y}^2,q)
           +B_{22}^{\mathcal{V}}(m_{x\mathcal{S}^{\prime}}^2,m_{\mathcal{S}^{\prime}y}^2,q)\Big)
\nonumber\\
&       + \Big(B_{21}^{\mathcal{V}}([m_{x\mathcal{S}}^2]^2,m_{\mathcal{S}y}^2,q)+B_{21}^{\mathcal{V}}([m_{\mathcal{S}y}^2]^2,m_{x\mathcal{S}}^2,q)\Big) \Big(
          + 32m_{x\mathcal{S}}^4\Big(2L^8_r-L^5_r\Big)
\nonumber \\&\hspace*{1cm}
           +  32m_{x\mathcal{S}}^2m_{\mathcal{S}^{\prime}\mathcal{S}^{\prime}}^2\Big(2L^6_r-L^4_r\Big)\Big)
\nonumber\\
&         - 16B_{22}^{\mathcal{V}}(m_{x\mathcal{S}}^2,m_{\mathcal{S}y}^2,q)L^9_{r}
          - B_2^{\mathcal{V}\alpha}(m_{x\mathcal{S}}^2,m_{\mathcal{S}y}^2,q)B_{2\alpha}^{\mathcal{V}}(m_{x\mathcal{S}^{\prime}}^2,m_{\mathcal{S}^{\prime}y}^2,q).
\end{align}

\begin{align}
  \label{eq:Pi1p6}
  F_0^2&\Pi_{1\pi^+_v}^{\mathcal{V}(6)} = \nonumber\\  &
\Big(A^{\mathcal{V}}(m^2_{xy}) - A^{\mathcal{V}}(m^2_{xx})\Big)
\Big( 8m^2_{xx}B_{22}^{\mathcal{V}}([m^2_{xx}]^2,m^2_{xy},q) - 2m^2_{xx}A^{\mathcal{V}}([m^2_{xx}]^2) + A^{\mathcal{V}}(m^2_{xy})
\nonumber\\&\hspace*{1cm}
 - A^{\mathcal{V}}(m^2_{xx}) \Big)
\nonumber\\
  &-m^2_{\mathcal{S}x}\Big(4B_{22}^{\mathcal{V}}(m^2_{x\mathcal{S}},[m^2_{\mathcal{S}y}]^2,q) + 4B_{22}^{\mathcal{V}}([m^2_{x\mathcal{S}}]^2,m^2_{\mathcal{S}y},q)- A^{\mathcal{V}}([m^2_{x\mathcal{S}}]^2) - A^{\mathcal{V}}([m^2_{y\mathcal{S}}]^2)\Big)
\nonumber\\  &
\hspace*{1cm}\times (16m^2_{\mathcal{S}^\prime \mathcal{S}^\prime}L_6^r - 8m^2_{\mathcal{S}^\prime \mathcal{S}^\prime}L_4^r + 16m^2_{\mathcal{S}x}L_8^r - 8m^2_{\mathcal{S}x}L_5^r + A^{\mathcal{V}}(\mathcal{D}_{x\mathcal{S}}))\nonumber\\
  &-2\Big(A^{\mathcal{V}}(m^2_{x\mathcal{S}}) + A^{\mathcal{V}}(m^2_{y\mathcal{S}})\Big)\Big(2L_{10}^rq^2+B_{22}^{\mathcal{V}}(m^2_{x\mathcal{S}^\prime},m^2_{\mathcal{S}^\prime y},q)\Big)\nonumber\\
  &+q^\beta \Big(A_\beta^{\mathcal{V}}(m^2_{x\mathcal{S}}) - A_\beta^{\mathcal{V}}(m^2_{\mathcal{S}^\prime \mathcal{S}})\Big)\Big(4B_{32}^{\mathcal{V}}([m^2_{x\mathcal{S}^\prime}]^2,m^2_{\mathcal{S}^\prime y},q)\Big)\nonumber\\
  &-q^\beta\Big(A_\beta^{\mathcal{V}}(m^2_{y\mathcal{S}}) - A_\beta^{\mathcal{V}}(m^2_{\mathcal{S}^\prime \mathcal{S}})\Big)\Big(4B_{22}^{\mathcal{V}}(m^2_{x\mathcal{S}^\prime},[m^2_{\mathcal{S}^\prime y}]^2,q) - 4B_{32}^{\mathcal{V}}(m^2_{x\mathcal{S}^\prime},[m^2_{\mathcal{S}^\prime y}]^2,q)\Big)\nonumber\\
  &- 2B_{22}^{\mathcal{V}}(m^2_{x\mathcal{S}},m^2_{\mathcal{S}y},q) \Big(8L_9^rq^2 - 2B_{22}^{\mathcal{V}}(m^2_{x\mathcal{S}^\prime},m^2_{\mathcal{S}^\prime y},q)\Big)
 +8L_9^rq^\beta\Big(A_\beta^{\mathcal{V}}(m^2_{x\mathcal{S}}) - A_\beta^{\mathcal{V}}(m^2_{y\mathcal{S}})\Big)
\nonumber\\ &
 + A^{\mathcal{V}}(m^2_{y\mathcal{S}})A^{\mathcal{V}}(m^2_{x\mathcal{S}^\prime})
+ A_{}^{\mathcal{V}\beta}([m^2_{x\mathcal{S}^\prime}]^2)\Big(A_{\beta}^{\mathcal{V}}(m^2_{\mathcal{S}^\prime \mathcal{S}}) - A_{\beta}^{\mathcal{V}}(m^2_{x\mathcal{S}})\Big)
\nonumber\\
  &+ A_{}^{\mathcal{V}\beta}([m^2_{y\mathcal{S}^\prime}]^2)\Big(A_{\beta}^{\mathcal{V}}(m^2_{\mathcal{S}^\prime \mathcal{S}}) - A_{\beta}^{\mathcal{V}}(m^2_{y\mathcal{S}})\Big).
\end{align}

\begin{align}
\label{eq:Pi2p6}
  F_0^2&\Pi^{\mathcal{V}(6)\mu\nu}_{2\pi^+_v} = \nonumber\\
&         + A^{\mathcal{V}\rho}(m_{x\mathcal{S}}^2)  \Big(
          + 2B_{22}^{\mathcal{V}}([m_{x\mathcal{S}^{\prime}}^2]^2,m_{\mathcal{S}^{\prime}y}^2,q)\delta_{\rho}^{\nu}q^{\mu}
          - 4B_32^{\mathcal{V}}([m_{x\mathcal{S}^{\prime}}^2]^2,m_{\mathcal{S}^{\prime}y}^2,q)\delta_{\rho}^{\nu}q^{\mu}
\nonumber\\
&\hspace*{1cm}          - \frac{1}{3}B^{\mathcal{V}}(m_{x\mathcal{S}^{\prime}}^2,m_{\mathcal{S}^{\prime}y}^2,q)\delta_{\rho}^{\nu}q^{\mu}
          + \frac{2}{3}B_1^{\mathcal{V}}(m_{x\mathcal{S}^{\prime}}^2,m_{\mathcal{S}^{\prime}y}^2,q)\delta_{\rho}^{\nu}q^{\mu}
          - 2B_{33\rho}^{\mathcal{V}\mu\nu}([m_{x\mathcal{S}^{\prime}}^2]^2,m_{\mathcal{S}^{\prime}y}^2,q)
\nonumber\\& \hspace*{1cm}
          + 4\delta_{\rho}^{\nu}L^9_{r}q^{\mu}
          +(\mu\leftrightarrow \nu)\Big)
\nonumber\\
&       + A^{\mathcal{V}\rho}(m_{y\mathcal{S}}^2)  \Big(
          - 2B_{22}^{\mathcal{V}}([m_{\mathcal{S}^{\prime}y}^2]^2,m_{x\mathcal{S}^{\prime}}^2,q)\delta_{\rho}^{\nu}q^{\mu}
          + 4B_32^{\mathcal{V}}([m_{\mathcal{S}^{\prime}y}^2]^2,m_{x\mathcal{S}^{\prime}}^2,q)\delta_{\rho}^{\nu}q^{\mu}
\nonumber\\
&\hspace*{1cm}     - \frac{2}{3}B^{\mathcal{V}}(m_{x\mathcal{S}^{\prime}}^2,m_{\mathcal{S}^{\prime}y}^2,q)\delta_{\rho}^{\nu}q^{\mu}
          + \frac{4}{3}B_1^{\mathcal{V}}(m_{x\mathcal{S}^{\prime}}^2,m_{\mathcal{S}^{\prime}y}^2,q)\delta_{\rho}^{\nu}q^{\mu}
          + 2B_{33\rho}^{\mathcal{V}\mu\nu}([m_{\mathcal{S}^{\prime}y}^2]^2,m_{x\mathcal{S}^{\prime}}^2,q)
\nonumber\\&\hspace*{1cm}
          - 4\delta_{\rho}^{\nu}L^9_{r}q^{\mu}
          +(\mu\leftrightarrow \nu)\Big)
\nonumber\\
&       + A^{\mathcal{V}\rho}(m_{\mathcal{S}\mathcal{S}^{\prime}}^2)  \Big(
         + 2B_{33\rho}^{\mathcal{V}\mu\nu}([m_{x\mathcal{S}}^2]^2,m_{\mathcal{S}y}^2,q)
          - 2B_{33\rho}^{\mathcal{V}\mu\nu}([m_{\mathcal{S}y}^2]^2,m_{x\mathcal{S}}^2,q)
\nonumber\\&\hspace*{1cm}
         + 4B_32^{\mathcal{V}}([m_{x\mathcal{S}}^2]^2,m_{\mathcal{S}y}^2,q)\delta_{\rho}^{\nu}q^{\mu}
          - 4B_32^{\mathcal{V}}([m_{\mathcal{S}y}^2]^2,m_{x\mathcal{S}}^2,q)\delta_{\rho}^{\nu}q^{\mu}
\nonumber\\&\hspace*{1cm}
         - 2B_{22}^{\mathcal{V}}([m_{x\mathcal{S}}^2]^2,m_{\mathcal{S}y}^2,q)\delta_{\rho}^{\nu}q^{\mu}
          + 2B_{22}^{\mathcal{V}}([m_{\mathcal{S}y}^2]^2,m_{x\mathcal{S}}^2,q)\delta_{\rho}^{\nu}q^{\mu}
         + B^{\mathcal{V}}(m_{x\mathcal{S}}^2,m_{\mathcal{S}y}^2,q)\delta_{\rho}^{\nu}q^{\mu}
\nonumber\\&\hspace*{1cm}
          - 2B_1^{\mathcal{V}}(m_{x\mathcal{S}}^2,m_{\mathcal{S}y}^2,q)\delta_{\rho}^{\nu}q^{\mu}
          +(\mu\leftrightarrow \nu)\Big)
\nonumber\\
&       + \frac{2}{3}\delta_{\alpha}^{\mu}\delta_{\beta}^{\nu}q^2B_2^{\mathcal{V}\alpha}(m_{x\mathcal{S}}^2,m_{\mathcal{S}y}^2,q)B_2^{\mathcal{V}\beta}(m_{x\mathcal{S}^{\prime}}^2,m_{\mathcal{S}^{\prime}y}^2,q)
\nonumber\\
&       + B_2^{\mathcal{V}\alpha}(m_{x\mathcal{S}}^2,m_{\mathcal{S}y}^2,q)  \Big(
          + \frac{4}{3}B_{21}^{\mathcal{V}}(m_{x\mathcal{S}^{\prime}}^2,m_{\mathcal{S}^{\prime}y}^2,q)\delta_{\alpha}^{\mu}q^{\nu}q^2
          + \frac{10}{3}B_{22}^{\mathcal{V}}(m_{x\mathcal{S}^{\prime}}^2,m_{\mathcal{S}^{\prime}y}^2,q)\delta_{\alpha}^{\mu}q^{\nu}
\nonumber\\
&\hspace*{1cm}          - \frac{2}{3}B^{\mathcal{V}}(m_{x\mathcal{S}^{\prime}}^2,m_{\mathcal{S}^{\prime}y}^2,q)\delta_{\alpha}^{\mu}q^{\nu}q^2
          + \frac{2}{3}B_1^{\mathcal{V}}(m_{x\mathcal{S}^{\prime}}^2,m_{\mathcal{S}^{\prime}y}^2,q)\delta_{\alpha}^{\mu}q^{\nu}q^2
          - \frac{5}{3}A^{\mathcal{V}}(m_{y\mathcal{S}^{\prime}}^2)\delta_{\alpha}^{\mu}q^{\nu}
\nonumber\\&\hspace*{1cm}
          + \frac{2}{3}A^{\mathcal{V}\rho}(m_{x\mathcal{S}^{\prime}}^2)\delta_{\alpha}^{\mu}\delta_{\rho}^{\nu}
          + \frac{4}{3}A^{\mathcal{V}\rho}(m_{y\mathcal{S}^{\prime}}^2)\delta_{\alpha}^{\mu}\delta_{\rho}^{\nu}
         + 2A^{\mathcal{V}\rho}(m_{\mathcal{S}^{\prime}\mathcal{S}}^2)\delta_{\alpha}^{\mu}\delta_{\rho}^{\nu}
          - 4\delta_{\alpha}^{\mu}L^9_{r}q^{\nu}q^2
          +(\mu\leftrightarrow\nu)\Big)
\nonumber\\
&       + \Big(B_2^{\mathcal{V}\alpha}([m_{x\mathcal{S}}^2]^2,m_{\mathcal{S}y}^2,q) + B_2^{\mathcal{V}\alpha}([m_{\mathcal{S}y}^2]^2,m_{x\mathcal{S}}^2,q)\Big)  \Big(
         - 2A^{\mathcal{V}}(\mathcal{D}_{x\mathcal{S}})\delta_{\alpha}^{\mu}m_{x\mathcal{S}}^2q^{\nu}
\nonumber\\&\hspace*{1cm}
          - 16\delta_{\alpha}^{\mu}q^{\nu}m_{x\mathcal{S}}^4\Big[2L^8_{r}-L^5_{r}\Big]
          - 16\delta_{\alpha}^{\mu}q^{\nu}m_{x\mathcal{S}}^4\Big[2L^6_{r}-L^4_{r}\Big]
          +(\mu\leftrightarrow\nu)\Big)
\nonumber\\
&       + B_2^{\mathcal{V}\alpha}(m_{xy}^2,m_{xx}^2,q)B_2^{\mathcal{V}\beta}(m_{xy}^2,m_{xx}^2,q)  \Big(
          - \frac{4}{3}\delta_{\alpha}^{\mu}\delta_{\beta}^{\nu}q^2
          + 8\delta_{\alpha}^{\mu}\delta_{\beta}^{\nu}m_{xx}^2
          \Big)
\nonumber\\
&       + B_2^{\mathcal{V}\alpha}(m_{xy}^2,m_{xx}^2,q)  \Big(
          + 2B^{\mathcal{V}}(m_{xy}^2,m_{xx}^2,q)\delta_{\alpha}^{\mu}q^{\nu}q^2
          - 4B^{\mathcal{V}}(m_{xy}^2,m_{xx}^2,q)\delta_{\alpha}^{\mu}m_{xx}^2q^{\nu}
\nonumber\\&\hspace*{1cm}
          - 4B_1^{\mathcal{V}}(m_{xy}^2,m_{xx}^2,q)\delta_{\alpha}^{\mu}q^{\nu}q^2
          + 8B_1^{\mathcal{V}}(m_{xy}^2,m_{xx}^2,q)\delta_{\alpha}^{\mu}m_{xx}^2q^{\nu}
          - \frac{8}{3}A^{\mathcal{V}}(m_{xy}^2)\delta_{\alpha}^{\mu}q^{\nu}
\nonumber\\&\hspace*{1cm}
          + \frac{8}{3}A^{\mathcal{V}}(m_{xx}^2)\delta_{\alpha}^{\mu}q^{\nu}
          +(\mu\leftrightarrow\nu)\Big)
\nonumber\\
&       + 4B_2^{\mathcal{V}\alpha}([m_{xx}^2]^2,m_{xy}^2,q)  \Big(
          + A^{\mathcal{V}}(m_{xy}^2)\delta_{\alpha}^{\mu}m_{xx}^2q^{\nu}
          - A^{\mathcal{V}}(m_{xx}^2)\delta_{\alpha}^{\mu}m_{xx}^2q^{\nu}
          +(\mu\leftrightarrow\nu)\Big)
\nonumber\\
&       + B_{23}^{\mathcal{V}\alpha\beta}(m_{x\mathcal{S}}^2,m_{\mathcal{S}y}^2,q)  \Big(
          - 4B_{21}^{\mathcal{V}}(m_{x\mathcal{S}^{\prime}}^2,m_{\mathcal{S}^{\prime}y}^2,q)\delta_{\beta}^{\nu}q_{\alpha}q^{\mu}
          - 4B_{22}^{\mathcal{V}}(m_{x\mathcal{S}^{\prime}}^2,m_{\mathcal{S}^{\prime}y}^2,q)\delta_{\alpha}^{\mu}\delta_{\beta}^{\nu}
\nonumber\\&\hspace*{1cm}
          - \frac{2}{3}B^{\mathcal{V}}(m_{x\mathcal{S}^{\prime}}^2,m_{\mathcal{S}^{\prime}y}^2,q)\delta_{\beta}^{\nu}q_{\alpha}q^{\mu}
          + \frac{10}{3}B_1^{\mathcal{V}}(m_{x\mathcal{S}^{\prime}}^2,m_{\mathcal{S}^{\prime}y}^2,q)\delta_{\beta}^{\nu}q_{\alpha}q^{\mu}
\nonumber\\&\hspace*{1cm}
         + A^{\mathcal{V}}(m_{x\mathcal{S}^{\prime}}^2)\delta_{\alpha}^{\mu}\delta_{\beta}^{\nu}
          + A^{\mathcal{V}}(m_{y\mathcal{S}^{\prime}}^2)\delta_{\alpha}^{\mu}\delta_{\beta}^{\nu}
          + 8\delta_{\alpha}^{\mu}\delta_{\beta}^{\nu}L^9_{r}q^2
          - 8\delta_{\beta}^{\nu}L^9_{r}q_{\alpha}q^{\mu}
\nonumber\\&\hspace*{1cm}
          + \frac{4}{3}\delta_{\beta}^{\nu}\delta_{\rho}^{\mu}B_2^{\mathcal{V}\rho}(m_{x\mathcal{S}^{\prime}}^2,m_{\mathcal{S}^{\prime}y}^2,q)q_{\alpha}
          + 2\delta_{\beta}^{\nu}B_2^{\mathcal{V}\alpha}(m_{x\mathcal{S}^{\prime}}^2,m_{\mathcal{S}^{\prime}y}^2,q)q^{\mu}
          + (\mu\leftrightarrow\nu)\Big)
\nonumber\\
&       + B_{23}^{\mathcal{V}\alpha\beta}([m_{x\mathcal{S}}^2]^2,m_{\mathcal{S}y}^2,q)  \Big(
         + 2A^{\mathcal{V}}(\mathcal{D}_{x\mathcal{S}})\delta_{\alpha}^{\mu}\delta_{\beta}^{\nu}m_{x\mathcal{S}}^2
          + 2A^{\mathcal{V}\alpha}(m_{x\mathcal{S}^{\prime}}^2)\delta_{\beta}^{\nu}q^{\mu}
          + 2A^{\mathcal{V}\alpha}(m_{\mathcal{S}^{\prime}\mathcal{S}}^2)\delta_{\beta}^{\nu}q^{\mu}
\nonumber\\&\hspace*{1cm}
         + 16m_{x\mathcal{S}}^4\Big[2L^8_r-L^5_r\Big]\delta_{\alpha}^{\mu}\delta_{\beta}^{\nu}
          + 16m_{x\mathcal{S}}^2m_{\mathcal{S}^{\prime}\mathcal{S}^{\prime}}^2\Big[2L^6_r-L^4_r\Big]\delta_{\alpha}^{\mu}\delta_{\beta}^{\nu}
          +(\mu\leftrightarrow\nu)\Big)
\nonumber\\
&       + B_{23}^{\mathcal{V}\alpha\beta}([m_{\mathcal{S}y}^2]^2,m_{x\mathcal{S}}^2,q)  \Big(
          + 2A^{\mathcal{V}}(\mathcal{D}_{x\mathcal{S}})\delta_{\alpha}^{\mu}\delta_{\beta}^{\nu}m_{x\mathcal{S}}^2
          - 2A^{\mathcal{V}\alpha}(m_{y\mathcal{S}^{\prime}}^2)\delta_{\beta}^{\nu}q^{\mu}
          - 2A^{\mathcal{V}\alpha}(m_{\mathcal{S}^{\prime}\mathcal{S}}^2)\delta_{\beta}^{\nu}q^{\mu}
\nonumber\\&\hspace*{1cm}
          + 16m_{x\mathcal{S}}^4\Big[2L^8_r-L^5_r\Big]\delta_{\alpha}^{\mu}\delta_{\beta}^{\nu}
          + 16m_{x\mathcal{S}}^2m_{\mathcal{S}^{\prime}\mathcal{S}^{\prime}}^2\Big[2L^6_r-L^4_r\Big]\delta_{\alpha}^{\mu}\delta_{\beta}^{\nu}
          +(\mu\leftrightarrow\nu)\Big)
\nonumber\\
&       + \Big(B_{23}^{\mathcal{V}\alpha\beta}(m_{xy}^2,m_{xx}^2,q) - B_{23}^{\mathcal{V}\alpha\beta}(m_{xx}^2,m_{xy}^2,q)\Big)  \Big(
          + \frac{2}{3}B^{\mathcal{V}}(m_{xy}^2,m_{xx}^2,q)\delta_{\alpha}^{\mu}q_{\beta}q^{\nu}
\nonumber\\&\hspace*{1cm}
          - \frac{4}{3}B_1^{\mathcal{V}}(m_{xy}^2,m_{xx}^2,q)\delta_{\alpha}^{\mu}q_{\beta}q^{\nu}
          - \frac{4}{3}\delta_{\alpha}^{\mu}\delta_{\rho}^{\nu}B_2^{\mathcal{V}\rho}(m_{xy}^2,m_{xx}^2,q)q_{\beta}
          +(\mu\leftrightarrow\nu)\Big)
\nonumber\\
&       + 8B_{23}^{\mathcal{V}\alpha\beta}([m_{xx}^2]^2,m_{xy}^2,q)\delta_{\alpha}^{\mu}\delta_{\beta}^{\nu}m_{xx}^2\Big(A^{\mathcal{V}}(m_{xx}^2)-A^{\mathcal{V}}(m_{xy}^2)\Big)
\nonumber\\
&- 4B_{23}^{\mathcal{V}\mu\rho}(m_{x\mathcal{S}^{\prime}}^2,m_{\mathcal{S}^{\prime}y}^2,q)B_{23\rho}^{\mathcal{V}\nu}(
m_{x\mathcal{S}}^2,m_{\mathcal{S}y}^2,q).
\end{align}

\subsection{$\Pi_{XY}^{\mathcal{V}\mu\nu}$ at $p^6$}
The results presented below are for the various components of
 $\Pi_{XY}^{\mathcal{V}(6)}$. These are the infinite and finite
 volume $p^6$ expressions for $\Pi_{XY}$. Note that the expressions are for
 the zero twist case. The partially twisted case is considerably longer
 and will be implemented numerically
 in \textsc{CHIRON} \cite{Bijnens:2014gsa,chiron}.

\begin{align}
  \label{eq:Pi0XYp6}
  F_0^2&\Pi_{0XY}^{\mathcal{V}(6)} = \nonumber\\
&       + \frac{1}{2}q^2B^{\mathcal{V}}(m_{X\mathcal{S}}^2,m_{X\mathcal{S}}^2,q)B^{\mathcal{V}}(m_{Y\mathcal{S}}^2,m_{Y\mathcal{S}}^2,q)
       -\frac{1}{4}q^2B^{\mathcal{V}}(m_{X\mathcal{S}}^2,m_{X\mathcal{S}}^2,q)B^{\mathcal{V}}(m_{XY}^2,m_{XY}^2,q)
\nonumber\\&
       + B^{\mathcal{V}}(m_{X\mathcal{S}}^2,m_{X\mathcal{S}}^2,q)  
\Big(
          + B_{21}^{\mathcal{V}}(m_{Y\mathcal{S}}^2,m_{Y\mathcal{S}}^2,q)q^2
          - B_{21}^{\mathcal{V}}(m_{XY}^2,m_{XY}^2,q)q^2
\nonumber\\&\hspace*{1cm}
          + B_{22}^{\mathcal{V}}(m_{Y\mathcal{S}}^2,m_{Y\mathcal{S}}^2,q)
          - B_{22}^{\mathcal{V}}(m_{XY}^2,m_{XY}^2,q)
          - \frac{3}{2}B_{1}^{\mathcal{V}}(m_{Y\mathcal{S}}^2,m_{Y\mathcal{S}}^2,q)q^2
\nonumber\\&\hspace*{1cm}
          + B_{1}^{\mathcal{V}}(m_{XY}^2,m_{XY}^2,q)q^2
          - \frac{1}{2}A^{\mathcal{V}}(m_{Y\mathcal{S}}^2)
           + \frac{1}{2}A^{\mathcal{V}}(m_{XY}^2)
          \Big)
\nonumber\\
&       -\frac{1}{4}q^2 B^{\mathcal{V}}(m_{Y\mathcal{S}}^2,m_{Y\mathcal{S}}^2,q)B^{\mathcal{V}}(m_{XY}^2,m_{XY}^2,q)
\nonumber\\
&       + B^{\mathcal{V}}(m_{Y\mathcal{S}}^2,m_{Y\mathcal{S}}^2,q)  
\Big(
          + B_{21}^{\mathcal{V}}(m_{X\mathcal{S}}^2,m_{X\mathcal{S}}^2,q)q^2
          - B_{21}^{\mathcal{V}}(m_{XY}^2,m_{XY}^2,q)q^2
\nonumber\\&\hspace*{1cm}
          + B_{22}^{\mathcal{V}}(m_{X\mathcal{S}}^2,m_{X\mathcal{S}}^2,q)
          - B_{22}^{\mathcal{V}}(m_{XY}^2,m_{XY}^2,q)
          - \frac{3}{2}B_{1}^{\mathcal{V}}(m_{X\mathcal{S}}^2,m_{X\mathcal{S}}^2,q)q^2
\nonumber\\&\hspace*{1cm}
          + B_{1}^{\mathcal{V}}(m_{XY}^2,m_{XY}^2,q)q^2
          - \frac{1}{2}A^{\mathcal{V}}(m_{X\mathcal{S}}^2)
          + \frac{1}{2}A^{\mathcal{V}}(m_{XY}^2)
          \Big)
\nonumber\\
&       + B^{\mathcal{V}}(m_{XY}^2,m_{XY}^2,q)  \Big(
          - 4L^9_{r}q^2
          - B_{21}^{\mathcal{V}}(m_{X\mathcal{S}}^2,m_{X\mathcal{S}}^2,q)q^2
          - B_{21}^{\mathcal{V}}(m_{Y\mathcal{S}}^2,m_{Y\mathcal{S}}^2,q)q^2
\nonumber\\&\hspace*{1cm}
          - B_{22}^{\mathcal{V}}(m_{X\mathcal{S}}^2,m_{X\mathcal{S}}^2,q)
          - B_{22}^{\mathcal{V}}(m_{Y\mathcal{S}}^2,m_{Y\mathcal{S}}^2,q)
          + B_{1}^{\mathcal{V}}(m_{X\mathcal{S}}^2,m_{X\mathcal{S}}^2,q)q^2
\nonumber\\&\hspace*{1cm}
          + B_{1}^{\mathcal{V}}(m_{Y\mathcal{S}}^2,m_{Y\mathcal{S}}^2,q)q^2
          + \frac{1}{2}A^{\mathcal{V}}(m_{X\mathcal{S}}^2)
          + \frac{1}{2}A^{\mathcal{V}}(m_{Y\mathcal{S}}^2)
          \Big)
\nonumber\\
&       + 2B^{\mathcal{V}}(m_{XY}^2,[m_{XY}^2]^2,q)  
\Big(
          - 16m_{XY}^4L^8_{r}
          + 8m_{XY}^4L^5_{r}
          - 16m_{XY}^2m_{\mathcal{S}\mathcal{S}}L^6_{r}
          + 8m_{XY}^2m_{\mathcal{S}\mathcal{S}}L^4_{r}
\nonumber\\&\hspace*{1cm}
          - A^{\mathcal{V}}(\mathcal{D}_{XY})m_{XY}^2
          \Big)
       + 3q^2B_{1}^{\mathcal{V}}(m_{X\mathcal{S}}^2,m_{X\mathcal{S}}^2,q)B_{1}^{\mathcal{V}}(m_{Y\mathcal{S}}^2,m_{Y\mathcal{S}}^2,q)
\nonumber\\
&       -2q^2 B_{1}^{\mathcal{V}}(m_{X\mathcal{S}}^2,m_{X\mathcal{S}}^2,q)B_{1}^{\mathcal{V}}(m_{XY}^2,m_{XY}^2,q) 
       -2q^2B_{1}^{\mathcal{V}}(m_{Y\mathcal{S}}^2,m_{Y\mathcal{S}}^2,q)B_{1}^{\mathcal{V}}(m_{XY}^2,m_{XY}^2,q)
\nonumber\\
&       + 8L^9_{r}q^2B_{1}^{\mathcal{V}}(m_{XY}^2,m_{XY}^2,q)
\nonumber\\
&       + 8B_{1}^{\mathcal{V}}(m_{XY}^2,[m_{XY}^2]^2,q)  \Big(
          + 16m_{XY}^4L^8_{r}
          - 8m_{XY}^4L^5_{r}
          + 16m_{XY}^2m_{\mathcal{S}\mathcal{S}}L^6_{r}
          - 8m_{XY}^2m_{\mathcal{S}\mathcal{S}}L^4_{r}
\nonumber\\&\hspace*{1cm}
          + A^{\mathcal{V}}(\mathcal{D}_{XY})m_{XY}^2
          \Big)
\nonumber\\
&       -4q^2 B_{21}^{\mathcal{V}}(m_{X\mathcal{S}}^2,m_{X\mathcal{S}}^2,q)B_{21}^{\mathcal{V}}(m_{Y\mathcal{S}}^2,m_{Y\mathcal{S}}^2,q)
       +4q^2 B_{21}^{\mathcal{V}}(m_{X\mathcal{S}}^2,m_{X\mathcal{S}}^2,q)B_{21}^{\mathcal{V}}(m_{XY}^2,m_{XY}^2,q) 
\nonumber\\
&       + B_{21}^{\mathcal{V}}(m_{X\mathcal{S}}^2,m_{X\mathcal{S}}^2,q)\Big(
          - 4B_{22}^{\mathcal{V}}(m_{Y\mathcal{S}}^2,m_{Y\mathcal{S}}^2,q)
          + 4B_{22}^{\mathcal{V}}(m_{XY}^2,m_{XY}^2,q)
          + 2A^{\mathcal{V}}(m_{Y\mathcal{S}}^2)
\nonumber\\&\hspace*{1cm}
          - 2A^{\mathcal{V}}(m_{XY}^2)
          \Big)
\nonumber\\
&       + 4q^2 B_{21}^{\mathcal{V}}(m_{Y\mathcal{S}}^2,m_{Y\mathcal{S}}^2,q)B_{21}^{\mathcal{V}}(m_{XY}^2,m_{XY}^2,q) 
\nonumber\\
&       + B_{21}^{\mathcal{V}}(m_{Y\mathcal{S}}^2,m_{Y\mathcal{S}}^2,q)\Big(
          - 4B_{22}^{\mathcal{V}}(m_{X\mathcal{S}}^2,m_{X\mathcal{S}}^2,q)
          + 4B_{22}^{\mathcal{V}}(m_{XY}^2,m_{XY}^2,q)
          + 2A^{\mathcal{V}}(m_{X\mathcal{S}}^2)
\nonumber\\&\hspace*{1cm}
          - 2A^{\mathcal{V}}(m_{XY}^2)
          \Big)
\nonumber\\
&       + B_{21}^{\mathcal{V}}(m_{XY}^2,m_{XY}^2,q)  \Big(
          + 4B_{22}^{\mathcal{V}}(m_{X\mathcal{S}}^2,m_{X\mathcal{S}}^2,q)
          + 4B_{22}^{\mathcal{V}}(m_{Y\mathcal{S}}^2,m_{Y\mathcal{S}}^2,q)
          - 2A^{\mathcal{V}}(m_{X\mathcal{S}}^2)
\nonumber\\&\hspace*{1cm}
          - 2A^{\mathcal{V}}(m_{Y\mathcal{S}}^2)
          \Big)
\nonumber\\
&       + 8B_{21}^{\mathcal{V}}(m_{XY}^2,[m_{XY}^2]^2,q)  \Big(
          - 16m_{XY}^4L^8_{r}
          + 8m_{XY}^4L^5_{r}
          - 16m_{XY}^2m_{\mathcal{S}\mathcal{S}}L^6_{r}
          + 8m_{XY}^2m_{\mathcal{S}\mathcal{S}}L^4_{r}
\nonumber\\&\hspace*{1cm}
          - A^{\mathcal{V}}(\mathcal{D}_{XY})m_{XY}^2
          \Big)
\nonumber\\&
       + 16B_{22}^{\mathcal{V}}(m_{XY}^2,m_{XY}^2,q)L^9_{r}
          + 8A^{\mathcal{V}}(m_{XY}^2)L^{10}_{r}.
\end{align}

\begin{align}
  \label{eq:Pi1XYp6}
  F_0^2&\Pi_{1XY}^{\mathcal{V}(6)} = 
+8m^2_{XY}(B_{22}^{\mathcal{V}}(m^2_{XY},[m^2_{XY}]^2,q)
\nonumber\\&
 - \frac{1}{4}A^{\mathcal{V}}([m^2_{XY}]^2))\Big(
  16m^2_{\mathcal{S}\mathcal{S}}L_6^r - 8m^2_{\mathcal{S}\mathcal{S}}L_4^r + A^{\mathcal{V}}(\mathcal{D}_{XY}) + 16m^2_{XY}L_8^r - 8m^2_{XY}L_5^r
  \Big)\nonumber\\
  &+2\Big(A^{\mathcal{V}}(m^2_{XY}) - A^{\mathcal{V}}(m^2_{X\mathcal{S}})\Big) B_{22}^{\mathcal{V}}(m^2_{Y\mathcal{S}},m^2_{Y\mathcal{S}},q)\nonumber\\
  &+ 2\Big(A^{\mathcal{V}}(m^2_{XY}) - A^{\mathcal{V}}(m^2_{Y\mathcal{S}})\Big) B_{22}^{\mathcal{V}}(m^2_{X\mathcal{S}},m^2_{X\mathcal{S}},q)\nonumber\\
  &+\Big(A^{\mathcal{V}}(m^2_{X\mathcal{S}}) + A^{\mathcal{V}}(m^2_{Y\mathcal{S}})\Big) \Big( 2B_{22}^{\mathcal{V}}(m^2_{XY},m^2_{XY},q)- A^{\mathcal{V}}(m^2_{XY})\Big)\nonumber\\
  &-4\Big(B_{22}^{\mathcal{V}}(m^2_{X\mathcal{S}},m^2_{X\mathcal{S}},q)B_{22}^{\mathcal{V}}(m^2_{XY},m^2_{XY},q) + B_{22}^{\mathcal{V}}(m^2_{Y\mathcal{S}},m^2_{Y\mathcal{S}},q)B_{22}^{\mathcal{V}}(m^2_{XY},m^2_{XY},q)\nonumber\\  &
\hspace*{1cm}- B_{22}^{\mathcal{V}}(m^2_{Y\mathcal{S}},m^2_{Y\mathcal{S}},q)B_{22}^{\mathcal{V}}(m^2_{X\mathcal{S}},m^2_{X\mathcal{S}},q)\Big)
\nonumber\\  &
+ A^{\mathcal{V}}(m^2_{Y\mathcal{S}})A^{\mathcal{V}}(m^2_{X\mathcal{S}})
+ 16B_{22}^{\mathcal{V}}(m^2_{XY},m^2_{XY},q)L_9^r q^2 + 8A^{\mathcal{V}}(m^2_{XY})L_{10}^rq^2.
\end{align}

\begin{align}
  \label{eq:Pi2XYp6}
  F_0^2&\Pi^{\mathcal{V}(6)\mu\nu}_{2XY} = \nonumber\\ &
       + B_{23}^{\mathcal{V}\alpha\beta}(m_{X\mathcal{S}}^2,m_{X\mathcal{S}}^2,q)  
\Big(
          - 4B_{21}^{\mathcal{V}}(m_{Y\mathcal{S}}^2,m_{Y\mathcal{S}}^2,q)\delta_{\beta}^{\mu}q_{\alpha}q^{\nu}
          + 4B_{21}^{\mathcal{V}}(m_{XY}^2,m_{XY}^2,q)\delta_{\beta}^{\mu}q_{\alpha}q^{\nu}
\nonumber\\&\hspace*{1cm}
          - 4B_{22}^{\mathcal{V}}(m_{Y\mathcal{S}}^2,m_{Y\mathcal{S}}^2,q)\delta_{\alpha}^{\mu}\delta_{\beta}^{\nu}
          + 4B_{22}^{\mathcal{V}}(m_{XY}^2,m_{XY}^2,q)\delta_{\alpha}^{\mu}\delta_{\beta}^{\nu}
          + B^{\mathcal{V}}(m_{Y\mathcal{S}}^2,m_{Y\mathcal{S}}^2,q)\delta_{\beta}^{\mu}q_{\alpha}q^{\nu}
\nonumber\\&\hspace*{1cm}
          - B^{\mathcal{V}}(m_{XY}^2,m_{XY}^2,q)\delta_{\beta}^{\mu}q_{\alpha}q^{\nu}
          + 2A^{\mathcal{V}}(m_{Y\mathcal{S}}^2)\delta_{\alpha}^{\mu}\delta_{\beta}^{\nu}
          - 2A^{\mathcal{V}}(m_{XY}^2)\delta_{\alpha}^{\mu}\delta_{\beta}^{\nu}
          \Big)
\nonumber\\
&       + B_{23}^{\mathcal{V}\alpha\beta}(m_{Y\mathcal{S}}^2,m_{Y\mathcal{S}}^2,q)  
\Big(
          - 4B_{21}^{\mathcal{V}}(m_{X\mathcal{S}}^2,m_{X\mathcal{S}}^2,q)\delta_{\beta}^{\nu}q_{\alpha}q^{\mu}
          + 4B_{21}^{\mathcal{V}}(m_{XY}^2,m_{XY}^2,q)\delta_{\beta}^{\nu}q_{\alpha}q^{\mu}
\nonumber\\&\hspace*{1cm}
          - 4B_{22}^{\mathcal{V}}(m_{X\mathcal{S}}^2,m_{X\mathcal{S}}^2,q)\delta_{\alpha}^{\mu}\delta_{\beta}^{\nu}
          + 4B_{22}^{\mathcal{V}}(m_{XY}^2,m_{XY}^2,q)\delta_{\alpha}^{\mu}\delta_{\beta}^{\nu}
          + B^{\mathcal{V}}(m_{X\mathcal{S}}^2,m_{X\mathcal{S}}^2,q)\delta_{\beta}^{\nu}q_{\alpha}q^{\mu}
\nonumber\\&\hspace*{1cm}
          - B^{\mathcal{V}}(m_{XY}^2,m_{XY}^2,q)\delta_{\beta}^{\nu}q_{\alpha}q^{\mu}
          + 2A^{\mathcal{V}}(m_{X\mathcal{S}}^2)\delta_{\alpha}^{\mu}\delta_{\beta}^{\nu}
          - 2A^{\mathcal{V}}(m_{XY}^2)\delta_{\alpha}^{\mu}\delta_{\beta}^{\nu}
          \Big)
\nonumber\\
&       + B_{23}^{\mathcal{V}\alpha\beta}(m_{XY}^2,m_{XY}^2,q)  
\Big(
          + 4B_{21}^{\mathcal{V}}(m_{X\mathcal{S}}^2,m_{X\mathcal{S}}^2,q)\delta_{\beta}^{\nu}q_{\alpha}q^{\mu}
          + 4B_{21}^{\mathcal{V}}(m_{Y\mathcal{S}}^2,m_{Y\mathcal{S}}^2,q)\delta_{\beta}^{\mu}q_{\alpha}q^{\nu}
\nonumber\\&\hspace*{1cm}
          + 4B_{22}^{\mathcal{V}}(m_{X\mathcal{S}}^2,m_{X\mathcal{S}}^2,q)\delta_{\alpha}^{\mu}\delta_{\beta}^{\nu}
          + 4B_{22}^{\mathcal{V}}(m_{Y\mathcal{S}}^2,m_{Y\mathcal{S}}^2,q)\delta_{\alpha}^{\mu}\delta_{\beta}^{\nu}
          - B^{\mathcal{V}}(m_{X\mathcal{S}}^2,m_{X\mathcal{S}}^2,q)\delta_{\beta}^{\nu}q_{\alpha}q^{\mu}
\nonumber\\&\hspace*{1cm}
          - B^{\mathcal{V}}(m_{Y\mathcal{S}}^2,m_{Y\mathcal{S}}^2,q)\delta_{\beta}^{\mu}q_{\alpha}q^{\nu}
          - 2A^{\mathcal{V}}(m_{X\mathcal{S}}^2)\delta_{\alpha}^{\mu}\delta_{\beta}^{\nu}
          - 2A^{\mathcal{V}}(m_{Y\mathcal{S}}^2)\delta_{\alpha}^{\mu}\delta_{\beta}^{\nu}
          - 16\delta_{\alpha}^{\mu}\delta_{\beta}^{\nu}L^9_{r}q^2
\nonumber\\&\hspace*{1cm}
          + 8\delta_{\beta}^{\nu}L^9_{r}q_{\alpha}q^{\mu}
          + 8\delta_{\beta}^{\mu}L^9_{r}q_{\alpha}q^{\nu}
          \Big)
\nonumber\\
&       + 8B_{23}^{\mathcal{V}\alpha\beta}(m_{XY}^2,[m_{XY}^2]^2,q)  
\Big(
          - A^{\mathcal{V}}(\mathcal{D}_{XY})\delta_{\alpha}^{\mu}\delta_{\beta}^{\nu}m_{XY}^2
          - 16\delta_{\alpha}^{\mu}\delta_{\beta}^{\nu}m_{XY}^4L^8_{r}
          + 8\delta_{\alpha}^{\mu}\delta_{\beta}^{\nu}m_{XY}^4L^5_{r}
\nonumber\\&\hspace*{1cm}
          - 16\delta_{\alpha}^{\mu}\delta_{\beta}^{\nu}m_{XY}^2m_{\mathcal{S}\mathcal{S}}L^6_{r}
          + 8\delta_{\alpha}^{\mu}\delta_{\beta}^{\nu}m_{XY}^2m_{\mathcal{S}\mathcal{S}}L^4_{r}
          \Big)
\nonumber\\
&       - 4\Big(B_{23}^{\mathcal{V}\alpha\nu}(m_{Y\mathcal{S}}^2,m_{Y\mathcal{S}}^2,q)B_{23}^{\mathcal{V}\alpha\mu}(m_{X\mathcal{S}}^2,m_{X\mathcal{S}}^2,q)
          - B_{23}^{\mathcal{V}\alpha\nu}(m_{Y\mathcal{S}}^2,m_{Y\mathcal{S}}^2,q)B_{23}^{\mathcal{V}\alpha\mu}(m_{XY}^2,m_{XY}^2,q)
\nonumber\\&\hspace*{1cm}
          - B_{23}^{\mathcal{V}\alpha\nu}(m_{XY}^2,m_{XY}^2,q)B_{23}^{\mathcal{V}\alpha\mu}(m_{X\mathcal{S}}^2,m_{X\mathcal{S}}^2,q)\Big).
\end{align}

\subsection{$\left<\bar q \gamma^\mu q\right>^{\mathcal{V}}$ at $p^6$}

\begin{align}
\label{eq:VEVp6}
  \left<\bar q \gamma^\mu q\right>^{\mathcal{V}(6)} = \hskip-1.8cm&
\nonumber\\  &
+ A^{\mathcal{V}\mu}(m_{q\mathcal{S}}^2)\Big(A^{\mathcal{V}}(m_{q\mathcal{S}^\prime}^2)-2A_{22}^{\mathcal{V}}([m_{q\mathcal{S}^\prime}^2]^2)\Big)
\nonumber\\&
  + 2m_{q\mathcal{S}}^2A^{\mathcal{V}\mu}([m_{q\mathcal{S}}^2]^2)  \Big(
   16m_{\mathcal{S}^\prime\mathcal{S}^\prime}L_6^r
  - 8m_{\mathcal{S}^\prime\mathcal{S}^\prime}L_4^r
  +  A^{\mathcal{V}}(\mathcal{D}_{q\mathcal{S}})
  + 16m_{q\mathcal{S}}^2L_8^r
  - 8m_{q\mathcal{S}}^2L_5^r
  \Big)
\nonumber\\
  &- A^{\mathcal{V}\mu}(m_{\mathcal{S}\mathcal{S}^\prime})\Big(A^{\mathcal{V}}(m_{q\mathcal{S}}^2) - 2 A_{22}^{\mathcal{V}}([m_{q\mathcal{S}}^2]^2)\Big)
  -2 A_{23}^{\mathcal{V}\beta\mu}([m_{q\mathcal{S}^\prime}^2]^2)\left(A^{\mathcal{V}}_{\beta}(m_{q\mathcal{S}}^2)-A^{\mathcal{V}}_{\beta}(m_{\mathcal{S}^\prime\mathcal{S}})\right)\,.
\end{align}

\end{document}